\documentclass[aps,prb,showpacs,a4paper,superscriptaddress,groupedaddress]{revtex4}
\usepackage[english]{babel}
\usepackage[utf8]{inputenc}
\usepackage{amsmath}
\usepackage{graphicx}
\usepackage[toc,page]{appendix}
\usepackage{multirow}
\usepackage{nonfloat}
\usepackage[colorinlistoftodos]{todonotes}
\usepackage{verbatim}

\begin{document}
\title{Ultrafast pulse phase shifts in a charged quantum dot - micropillar system}
\date{\today}

\author{G. Slavcheva}\email{g.slavcheva@jku.at}
\affiliation{Institute of Semiconductor and Solid State Physics, Johannes Kepler University Linz, Altenbergerstrasse 69, A-4040 Linz, Austria}
\author{M. Koleva}
\affiliation{Faculty of Life Sciences and Medicine, King's College London, SE1 1UL, United Kingdom}
\author{A. Rastelli}
\affiliation{Institute of Semiconductor and Solid State Physics, Johannes Kepler University Linz, Altenbergerstrasse 69, A-4040 Linz, Austria}

\begin{abstract}
We employ a quantum master equations approach based on a vectorial Maxwell-pseudospin model to compute the quantum evolution of the spin populations and coherences in the fundamental singlet trion transition of a negatively charged quantum dot embedded in a micropillar cavity. Excitation of the system is achieved through an ultrashort, either circularly or linearly polarised resonant pulse. By implementing a realistic micropillar cavity geometry, we numerically demonstrate a giant optical phase shift ($\sim \pm \pi/2$) of a resonant circularly polarised pulse in the weak-coupling regime. The phase shift that we predict considerably exceeds the experimentally observed Kerr rotation angle $(\sim{6 ^{\circ}})$ under a continuous-wave, linearly polarised excitation. By contrast, we show that a linearly polarised pulse is rotated to a much lesser extent of a few degrees. Depending on the initial boundary conditions, this is due to either retardation or advancement in the amplitude build-up in time of the orthogonal electric field component. Unlike previous published work, the dominant spin relaxation and decoherence processes are fully accounted for in the system dynamics. Our dynamical model can be used for optimisation of the optical polarisation rotation angle for realisation of spin-photon entanglement and ultrafast polarisation switching on a chip.
\end{abstract}
\maketitle
\section{Introduction}
\label{sec1}
Realisation of a deterministic spin-photon entanglement in semiconductor quantum dots (QDs) is one of the major goals on the path to developing integrated quantum photonics based on this platform. In general, there are two possible ways of realising a quantum superposition of states: through controlled rotations of either 1) the material spin qubit or 2) the photon polarisation state representing the qubit. One promising method of achieving a quantum superposition of matter states is selectively addressing individual charge-carrier spins using circularly polarised photons generated by an external source, and manipulating the spins through optically excited states (charged excitons) by employing the techniques of coherent quantum control and optical orientation \cite{Imamoglu2012}$^-$\cite{Vuckovic2016}. An alternative approach is to prepare high-fidelity photon polarisation states by making use of the optical polarisation rotation induced by the single spin confined in a QD. The spin-induced photon polarisation rotation acts as a spin-photon entangler and effectively performs the function of a controlled-phase gate. The quantum logic with photonic qubits relies on a physical implementation of such controlled-phase gates \cite{Fushman2008}.

Coupling a stationary spin qubit to an optical cavity leads to an enhanced efficiency of spin-photon interaction. Due to multiple reflections and round-trips between the mirrors, the photon polarisation rotation angle of the pulse accumulates appreciably and increases by several orders of magnitude \cite{KavokinPRB97}$^{,}$\cite{HuPRB2008}. A single QD strongly coupled to a cavity mode is a solid-state analogue of an atom-cavity system in quantum optics. Such cavity-dot systems are characterized by extremely large optical nonlinearities as the photons circulating in a high-finesse cavity can interact strongly through their coupling with a single QD. In this respect, a spin in a QD can entangle two coincident photons in a photon-based quantum logic \cite{HuMunroRarityPRB2008}$^-$\cite{LeeLawPRA2006}.

As has been pointed out in \cite{Nielsen}, for realisation of useful quantum gates controlled phase shifts of $\pi$ are necessary. A phase shift up to $\pi/4$ has been demonstrated in a single QD strongly coupled to a photonic crystal cavity \cite{Fushman2008}. Recently, a macroscopic Kerr rotation of $6 ^{\circ}$ of the photon polarization has been observed using continuous-wave (CW) excitation in both the strong- \cite{ArnoldNatComm2015} and weak-coupling \cite{Ruth2016} regimes in a charged QD-micropillar system. The Kerr nonlinearity results from non-resonant optical excitation of the atom (QD)-cavity system \cite{KimblePRL2004}.

A maximum phase shift of $\pi$ is predicted by a number of theoretical works \cite{Hofmann2003}$^{,}$\cite{HuPRB2008}$^{,}$\cite{Carmichael2008} in the strong coupling regime, which may be reached even without a high-finesse cavity \cite{ZimofenPRL2008}.  The theoretical approaches used to describe the origin of the optical polarisation phase shift induced by a single spin in atomic systems and QDs are limited to the strong coupling regime and use a number of approximations and idealised cavities, represented by a cavity loss rate. For instance, the reflection coefficient of the QD-cavity system in \cite{HuPRB2008} is obtained in the stationary case, assuming that the negative trion resides mostly in its ground electron spin-up (down) state (Fig.~\ref{fig:dot_micropillar} (b)). Optical Bloch equations have been used in \cite{Hofmann2003} to describe the nonlinear response of a two-level system at resonance with a one-sided cavity with negligible cavity losses, representing a great simplification of the realistic dot-cavity system.
Furthermore, within a simplified time-scale separation ('three-stage') model proposed in \cite{SmirnovPRB2015}, only the ground state resident electron spin relaxation is taken into account within a Markovian description under the assumption that all incoherent processes are perturbative. The spin relaxation and decoherence (spin-depolarising) dynamics of the trion states is ignored under the assumption of their arguably relatively long time scales compared to the trion decay rate and photon decay rate. However, the excited trion states' hole spin-flip relaxation, due to phonon-assisted processes \cite{Greilich} occurs at much shorter time scales and should be given due consideration in the system dynamics. In addition, it has been argued \cite{SmirnovPRB2015} that the spin-depolarisation (decoherence) processes have little influence on the system dynamics. A full treatment of the spin relaxation and decoherence non-Markovian dynamics is thus necessary to describe the complex system dynamics when the above perturbative assumption, as we shall show below, is no longer valid.

In this work, we develop a dynamical model and investigate numerically the light-matter interaction in realistic micropillar cavity-dot distributed Bragg reflector (DBR) geometries. As has been pointed out in \cite{Fushman2008}, the cavity-embedded QD is a highly nonlinear system and cannot be well described by a pure Kerr medium. We show that the resonant nonlinearities, associated with the negative trion transition, as opposed to Kerr nonlinearities, result in much larger phase shifts. In particular, we show that a resonant coherent interaction of an ultrashort circularly polarised pulse with a discrete multi-level system, representing the negative trion singlet transition (Fig.~\ref{fig:dot_micropillar} (b)), results in a giant phase shift of $\pi$ even in the weak-coupling regime. Unlike earlier theories \cite{HuPRB2008}$^,$\cite{SmirnovPRB2015} and experiments \cite{ArnoldNatComm2015}$^,$\cite{Ruth2016}, considering coherent scattering and reflection coefficient from the cavity-QD system, we model the pulse transmission through the dot-micropillar structure in a Faraday rotation configuration. We consider coherent interactions of an electromagnetic wave tuned in resonance with the trion transition. The resonant nonlinear response of the active medium can then be described in terms of an 'atomic' susceptibility. The 'atomic' phase shift experienced by a light pulse propagating through a resonantly absorbing/amplifying medium is due to the resonant coherent interaction of the propagating light pulse with the active medium \cite{SiegmanLasers}. Such phase shifts could be measured by interferometric methods or by homodyne or heterodyne detection, interfering the cavity-transmitted photons with a reference beam of known amplitude and phase as in Ref. \onlinecite{Fushman2008}.

Our quantum master equations approach is based on self-consistent solution of Maxwell's curl equations coupled through the medium polarisation to the Liouville-von Neumann equations for the density matrix evolution of a multi-level quantum system in a real coherence vector representation \cite{SlavchevaPRB2008}$^{,}$\cite{SlavchevaKoleva2017}. This approach allows to model the dynamics beyond the two-level system approximation and use an equivalent four-level system of the negative trion ($X^{-}$) transition in a QD, thereby mapping all dipole-allowed optical transitions, as well as describe the inter-level population dynamics taking place. The equations set is solved directly in the time domain by the Finite-Difference Time-Domain (FDTD) method, thereby allowing implementation of macroscopic boundary conditions in a realistic cavity-dot structure. In addition, the FDTD method describes properly the coupling between the forward and backward propagating electromagnetic waves within the cavity. The perfectly transmitting boundary conditions permit calculation of the cavity loss in a realistic cavity which can be inferred from the cavity mode width of the transmission peak in the stop band of the numerically simulated transmission spectrum.

Large phase shifts, or equivalently - in the case of linear or elliptical polarisation - rotation angles, are highly desirable for fabrication of phase gates for optical quantum computing as they enable performing high-fidelity gate operations. Moreover, the large photon polarisation rotation angles resulting from an enhancement of photon-spin interactions in optical cavities open avenues for using charged QD-cavity structures as ultrafast polarisation switches exploiting the optical Faraday rotation effect. In this respect, it is worthwhile to develop theoretical and numerical techniques for optimisation of the controlled photon qubit phase shifts. With the present study we aim to develop a theoretical framework and numerical tools for optimisation of the phase shift produced by realistic cavity-dot structures and devices.

The paper is organized as follows: The theoretical model and its numerical implementation is described in Sec. \ref{sec2}. We set up the model parameters and the micropillar structure geometry in \ref{sec3}. We study the following experimentally realisable excitation scenarios and calculate the quantum system evolution upon: (i) circularly polarised and (ii) linearly polarised ultrashort optical pulse with initial spin population prepared (by e.g. optical pumping) entirely in one of the doubly-degenerate trion ground levels, or (iii) linearly polarised excitation of a system initially in thermal equilibrium with equally distributed between the ground levels spin-up and spin-down populations. In Sec. \ref{sssec:3.2} we lay out the method for calculation of the phase shift induced by the resonant system during the pulse propagation across the cavity. The polarisation rotation angle is calculated in each of the above cases and we show that a maximum of $\sim \pm \pi/2$ is achieved with a resonant ultashort circularly polarised pulse. By contrast, we demonstrate that a linearly polarised pulse leads to a shift between the resonances corresponding to the two orthogonal field components. The latter results in an effective decrease of the optical rotation angle, as the two orthogonal field components oscillate with different frequencies and interfere destructively. The special case of initial spin population in thermal equilibrium which does not require initial optical spin pumping is discussed. In this case two dipole allowed transitions are simultaneously excited and the phase shift curve of the orthogonal field component is displaced, however in an opposite direction to the previous case, leading to destructive interference and effectively reducing the pulse polarisation rotation angle.

\section{Theoretical model}
\label{sec2}
\begin{figure}
\resizebox{.9\textwidth}{!}{%
\includegraphics[height=4 cm]{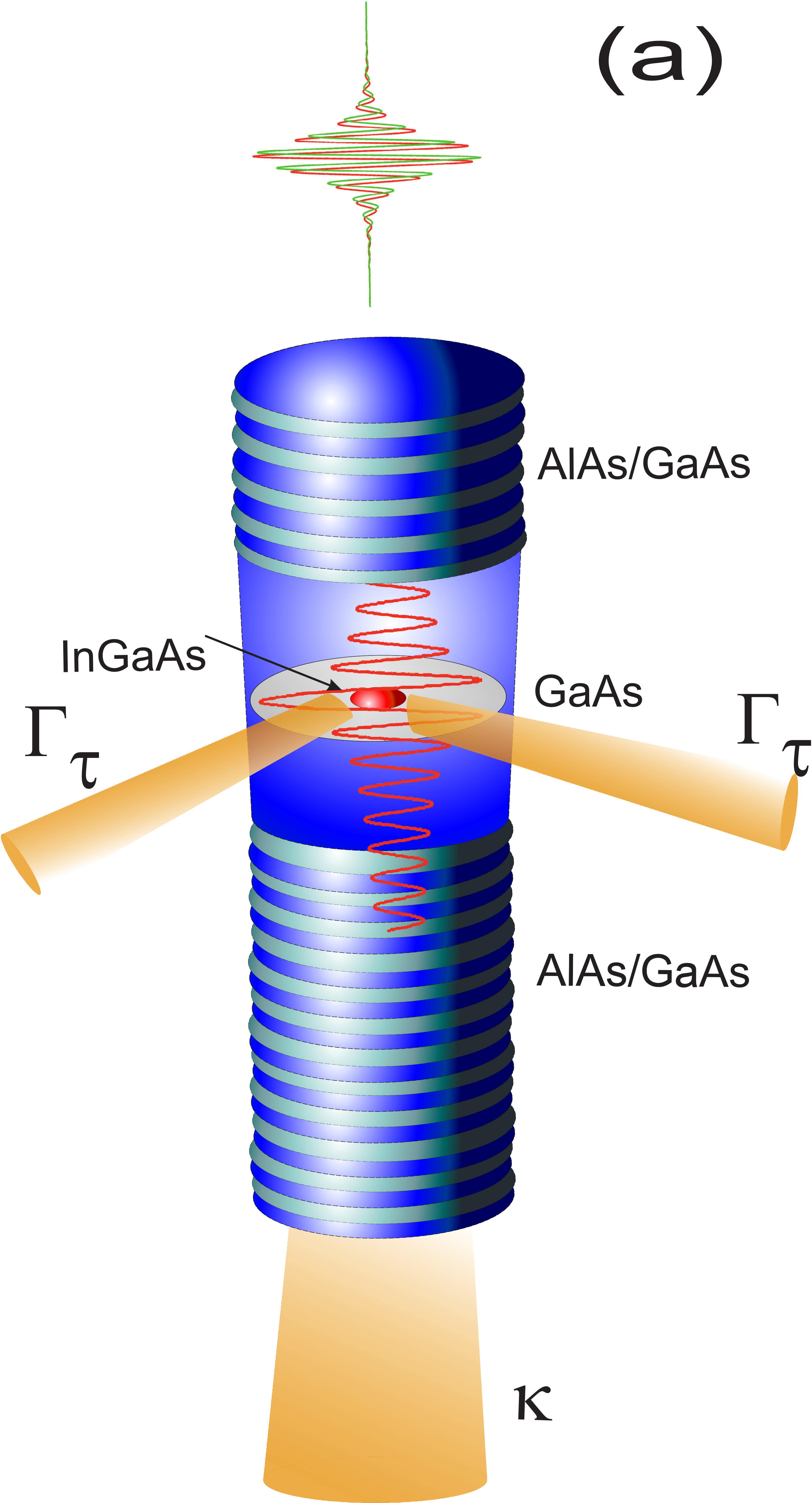}
\quad
\includegraphics[height=3 cm]{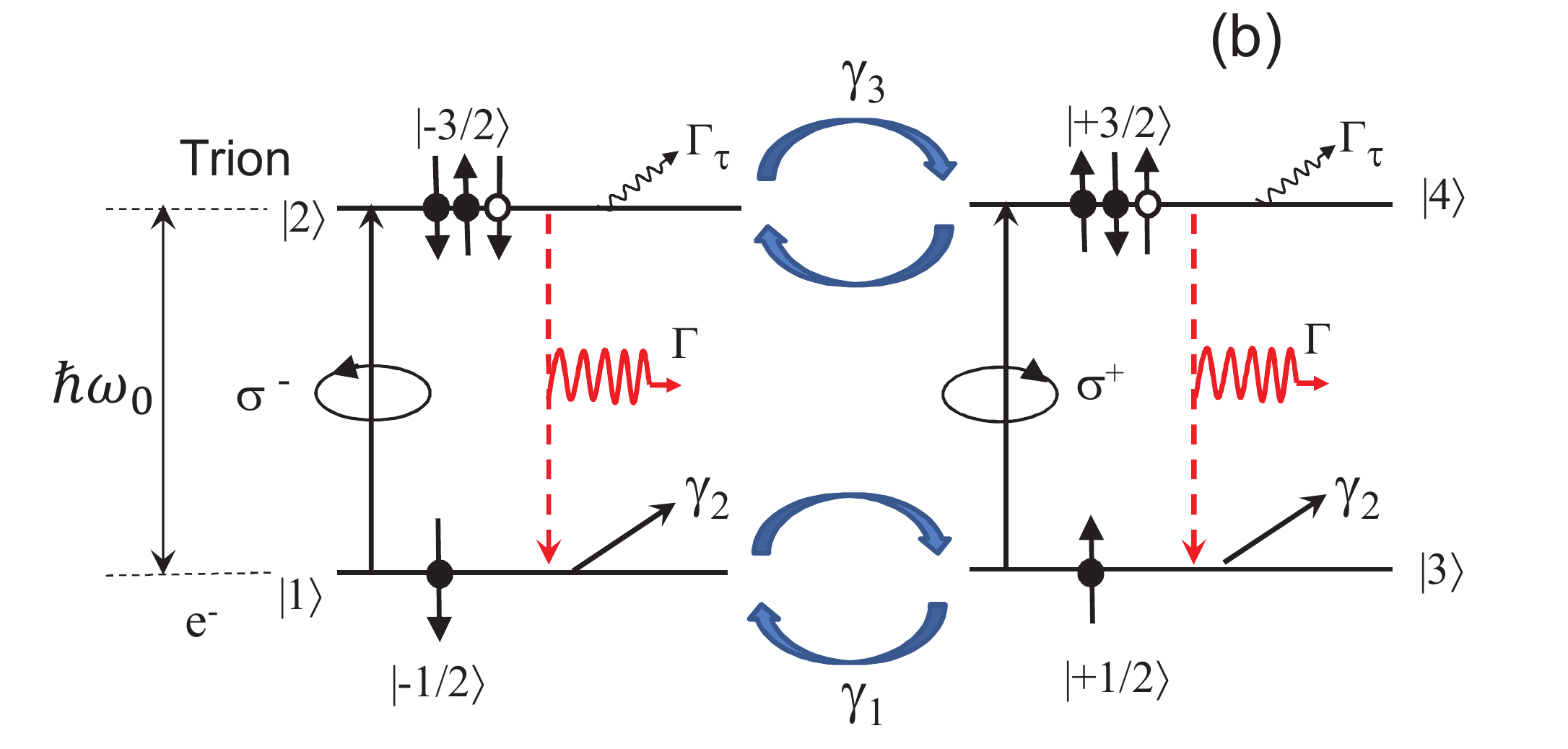}
}
\caption{(a) Sketch of a charged QD-micropillar structure; $\kappa$ - cavity mode decay rate through the mirrors, $\Gamma_{\tau}$ - trion decay rate to non-cavity modes. The micropillar cavity is pumped from the top (through an optical fibre) by an ultrashort circularly polarised pulse: $E_x$, $E_y$ field components shown in green/red, respectively. The DBR structure consists of $AlAs/GaAs$ alternating layers and the $InGaAs$ QD is positioned close to the antinode of the standing wave profile across the structure; (b) Energy-level scheme of a negative trion transition in a charged QD, resonantly-driven by either left- ($\sigma^{-}$) or right-($\sigma^{+}$) circularly polarised light (coherent transitions denoted by solid upward arrows). Solid and open circles denote electrons and holes. The levels are labeled by the total angular momentum projection ($J_z$) along the pulse propagation and QD growth $\textit{z}$-axis and the fundamental energy gap is $\hbar \omega_0$. Trion recombination (spontaneous emission) rate, $\Gamma$, is denoted by red dashed downward arrows; the curved blue arrows labelled $\gamma_1$, $\gamma_3$ denote the electron (due to hyperfine interaction with the nuclear spins in the host lattice) and hole spin-flip (due to phonon-assisted processes) rates, respectively; $\gamma_2$ is the electron spin-decoherence rate and $\Gamma_{\tau}=\gamma_{sp,nc}+\gamma_{dephasing}$ is the trion state decay rate, which consists of a combined contribution from spontaneous emission to non-cavity modes and spin-decoherence (dephasing) rate.
}
\label{fig:dot_micropillar}
\end{figure}

We consider a negatively charged InGaAs QD embedded in a micropillar cavity schematically shown in Fig.~\ref{fig:dot_micropillar}(a). The QD ground state is doubly-degenerate and  the single confined spin may be in either spin-down or spin-up state. Upon an optical excitation, an exciton is formed and the resident electron is promoted to an excited charged exciton three-particle doubly-degenerate state consisting of two electrons in a singlet state and a hole. An equivalent energy-level scheme of the fundamental trion singlet transition and the dipole-allowed optical transitions involving single photon excitation are displayed in Fig.~\ref{fig:dot_micropillar}(b).

The dynamical evolution of an open $N$ discrete-level quantum system under a time-dependent perturbation is described by the quantum master Liouville-von Neumann equation for the density operator, $\hat \rho $, in the Schr\"{o}dinger picture, modified to include damping in the system through longitudinal (population relaxation) ($\hat \sigma$) and transverse (decoherence) relaxation ($\hat \Gamma_t$) terms \cite{SlavchevaPRB2008}:
\begin{equation}
\frac{{\partial \hat \rho }}{{\partial t}} = \frac{i}{\hbar }\left[ {\hat \rho ,\hat H} \right] + \hat \sigma  - \hat \Gamma _t \cdot \hat \rho
\label{eq:Liouville_master}
\end{equation}
where $\hat H$ is the system Hamiltonian: $\hat H = \hat H_0  + \hat H_{{\mathop{\rm int}} } \left( t \right)$, with $\hat H_0$ being the unperturbed Hamiltonian of an N-level system: a diagonal matrix with eigenenergies, $\hbar \omega_k\,\ (k=1,...,\mathrm{N})$, of each level along the main diagonal, and $\hat H_{int}=-e {\bf \hat r}\cdot{\bf E}$ is a time-dependent dipole-coupling perturbation (not necessarily small), with $\bf \hat r = (\bf \hat x, \bf \hat y)$ being the local displacement operator.

In the specific case of a fundamental trion transition (no excited charged excitonic states are considered) in a charged QD, driven by an either left- or right elliptically (circularly when $E_x=E_y$)  polarised pulse with complex electric field vector, ${\bf E}\left( {{\bf r},t} \right) = E_x {\bf \hat e_x} \pm iE_y {\bf \hat e_y}$, with $\bf \hat e_x, \bf \hat e_y$ -- unit polarisation vectors, the system is described by $N=4$ levels (see Fig.~\ref{fig:dot_micropillar}(b)) and the explicit form of the system Hamiltonian is given by:
\begin{equation}
\hat H = \hbar \left( {\begin{array}{*{20}c}
   0 & { - \frac{1}{2}\left( {\Omega _x  - i\Omega _y } \right)} & 0 & 0  \\
   { - \frac{1}{2}\left( {\Omega _x  + i\Omega _y } \right)} & {\omega _0 } & 0 & 0  \\
   0 & 0 & 0 & { - \frac{1}{2}\left( {\Omega _x  + i\Omega _y } \right)}  \\
   0 & 0 & { - \frac{1}{2}\left( {\Omega _x  - i\Omega _y } \right)} & {\omega _0 }.  \\
\end{array}} \right)
\end{equation}
where we have defined the Rabi frequencies:  $\Omega _x  = \frac{\wp }{\hbar }E_x;\,\,\,\,\Omega _y  = \frac{\wp }{\hbar }E_y $ with ${\vec{\wp}} = \left\langle i \right.\left| {e{\bf \hat x}} \right|\left. j \right\rangle = \left\langle i \right.\left| {e{\bf \hat y}} \right|\left. j \right\rangle$ being the optical dipole moment matrix element between any pair of levels $|i\rangle$ and $|j\rangle$ (in this particular case $\wp  = \left| {\vec \wp } \right|= \left\langle 1 \right.\left| {e x} \right|\left. 2 \right\rangle = \left\langle 1 \right.\left| {e y} \right|\left. 2 \right\rangle =\left\langle 3 \right.\left| {e x} \right|\left. 4 \right\rangle = \left\langle 3 \right.\left| {e y} \right|\left. 4 \right\rangle$ along the electric field). Note that the anti-diagonal elements along the skew matrix diagonal ($H_{i,N+1-j},\,\,\ i,j=1,...,N$) in a realistic QD are not necessarily vanishing, as there may be hole mixing or a slight tilt of the quantization axis. These Hamiltonian elements, however, are important for spin pumping under external magnetic field. In this work we consider a zero magnetic field case ($B=0$) and thus the anti-diagonal elements are set to zero.

The dissipation in the system in (\ref{eq:Liouville_master}) is taken into account by separate contributions due to longitudinal spin relaxation processes, associated with spin population transfer between pair of levels, involving dipole-allowed transitions in the four-level system, and transverse spin decoherence processes involving transitions from a particular energy level, within the four-level system under consideration, to other external energy levels. The formalism used to describe the longitudinal and transverse spin relaxation processes in this particular system is outlined in the Appendix.

In order to take advantage of the $SU(N)$ Lie group dynamical symmetries, we have derived \cite{SlavchevaPRB2008}$^,$\cite{SlavchevaKoleva2017} equivalent master pseudospin equations of motion using real coherent state vector representation of the density matrix, which for the particular case of $N=4$ levels, read:
\begin{equation}
\frac{{\partial S_j }}{{\partial t}} = \left\{ \begin{array}{l}
 f_{jkl} \Gamma _k S_l  + \frac{1}{2}Tr\left( {\hat \sigma \cdot \hat \lambda _j } \right) - \frac{1}{{T_j }}\left( {S_j  - S_{je} } \right), \,\,\,\,\ j = 1,2,..,12 \\
 f_{jkl} \Gamma _k S_l  + \frac{1}{2}Tr\left( {\hat \sigma \cdot\hat \lambda _j } \right), \,\,\,\,\,\,\,\,\,\,\,\,\,\,\,\,\,\,\,\,\,\ j = 13,14,15, \\
 \end{array} \right.
 \label{eq:pseudospin_master}
\end{equation}
where $\Gamma_k$ is the torque vector, $\mathord{\buildrel{\lower3pt\hbox{$\scriptscriptstyle\frown$}}\over \lambda } _j$ are $4\times4$ matrices, known as generators of $SU(4)$ Lie group algebra. The $\lambda$-generators represent a generalisation of the Pauli matrices for the simplest two-level case to a system with an arbitrary number of discrete energy levels, $N$. We have defined in Eq. \ref{eq:pseudospin_master} an equilibrium coherence vector, $S_E=S_{1e}, S_{2e},...,S_{12e}$, and $T_j, \,\,\ j=1,2...,15$ are phenomenologically introduced nonuniform decay times describing the relaxation of the real state vector components towards their equilibrium values, $S_{je}$. The longitudinal spin population relaxation times are given by:
\begin{equation}
T_{13}  = \frac{4}{{2\Gamma _{21}  + \gamma _{13}  + \gamma _{24} }},\,\,\,T_{14}  = \frac{4}{{3\gamma _{13}  + \gamma _{24}  + 6\gamma _{31} }},\,\,T_{15}  = \frac{6}{{\gamma _{24}  + 3\left( {\gamma _{42}  + \Gamma _{43} } \right)}}
\end{equation}
where using the notations of Fig.~\ref{fig:dot_micropillar}(b) $\Gamma_{12}=\Gamma_{43}=\Gamma$ is the trion state spontaneous emission rate, $\gamma_{13}=\gamma_{31}=\gamma_1$ is the electron spin-flip relaxation rate between the lower-lying electron levels and $\gamma_{24}=\gamma_{42}=\gamma_3$ is the hole-spin flip population transfer rate between the upper-lying trion states. The transverse relaxation (spin decoherence) times are given by:$T_1  = T_2  = T_5  = T_6  = T_7  = T_8  = T_{11}  = T_{12}  = 1/\Gamma _\tau$ and $T_3  = T_4 \, = T_9  = T_{10}  = 1/\gamma _2$.

The vector Maxwell equations for a circularly polarized optical pulse exciting the trion transition in a four-level system, thereby inducing macroscopic dipole polarizations, $P_x$ and $P_y$ along the $x$- and $y$-directions respectively, in a plane perpendicular to the propagation direction, $z$, are given by:
\begin{eqnarray}
\begin{aligned}
& \frac{{\partial H_x \left( {z,t} \right)}}{{\partial t}} = \frac{1}{\mu }\frac{{\partial E_y \left( {z,t} \right)}}{{\partial z}} \\
& \frac{{\partial H_y \left( {z,t} \right)}}{{\partial t}} =  - \frac{1}{\mu }\frac{{\partial E_x \left( {z,t} \right)}}{{\partial z}} \\
& \frac{{\partial E_x \left( {z,t} \right)}}{{\partial t}} =  - \frac{1}{\varepsilon }\frac{{\partial H_y \left( {z,t} \right)}}{{\partial z}} - \frac{1}{\varepsilon }\frac{{\partial P_x \left( {z,t} \right)}}{{\partial t}} \\
& \frac{{\partial E_y \left( {z,t} \right)}}{{\partial t}} = \frac{1}{\varepsilon }\frac{{\partial H_x \left( {z,t} \right)}}{{\partial z}} - \frac{1}{\varepsilon }\frac{{\partial P_y \left( {z,t} \right)}}{{\partial t}}\,.
& \label{eq:Maxwell_vector_circ_polar}
\end{aligned}
\end{eqnarray}
The master pseudospin equations (\ref{eq:pseudospin_master}) are coupled to the Maxwell's equations (\ref{eq:Maxwell_vector_circ_polar}) through the medium polarisation for which the following relations have been derived for the polarisation components induced by a circularly polarised pulse\cite{SlavchevaPRB2008}:
\begin{equation}
\begin{array}{l}
 P_x  =  - \wp N_d S_1  \\
 P_y  =  - \wp N_d S_7  \\
\end{array}
\label{eq:polarisation}
\end{equation}
where $N_d$ is the resonant dipole density, or the number of resonantly excited four-level systems (charged QDs) per unit volume.

\begin{figure}
\resizebox{.9\textwidth}{!}{%
\includegraphics[height=3 cm]{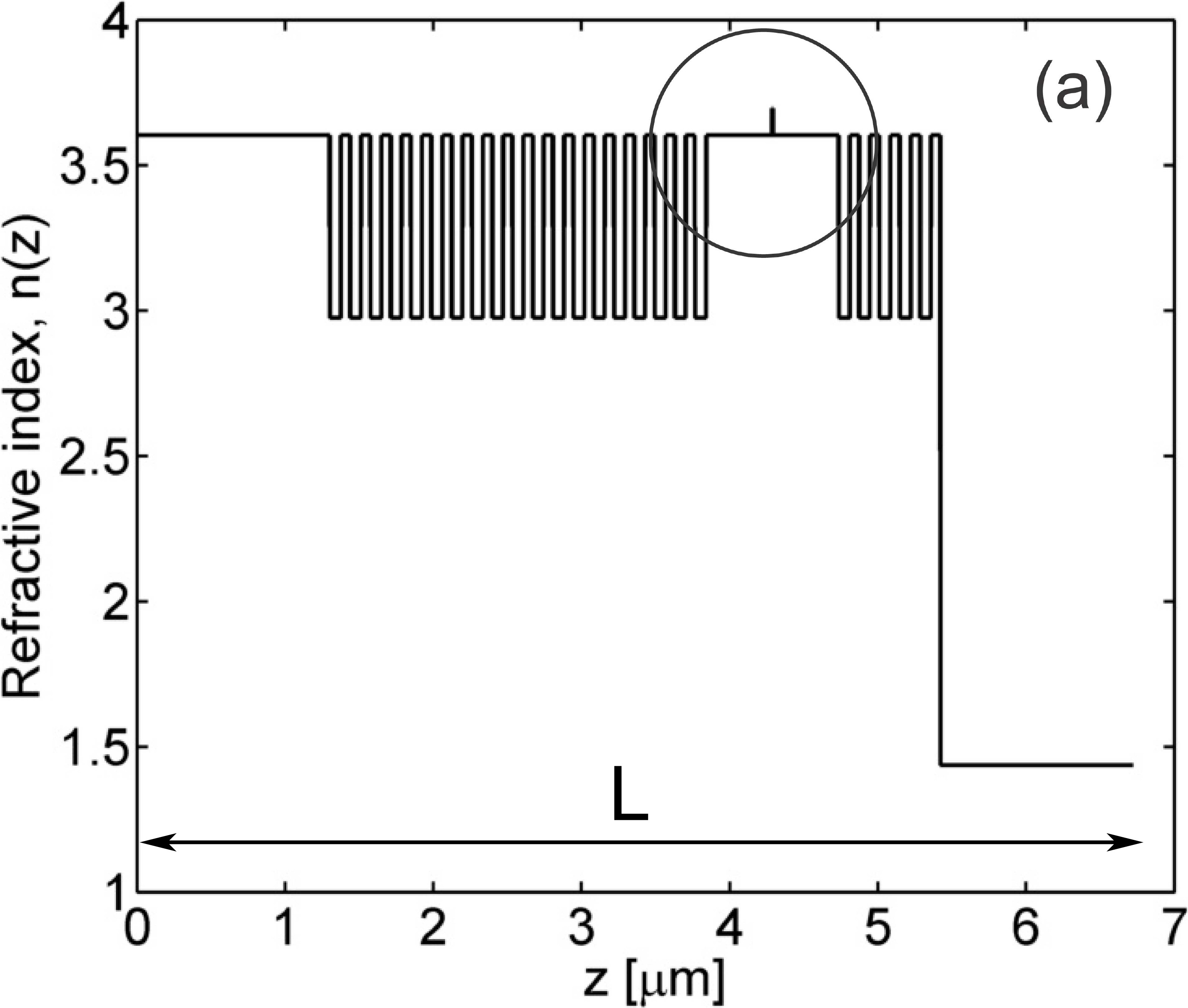}
\quad
\includegraphics[height=3 cm]{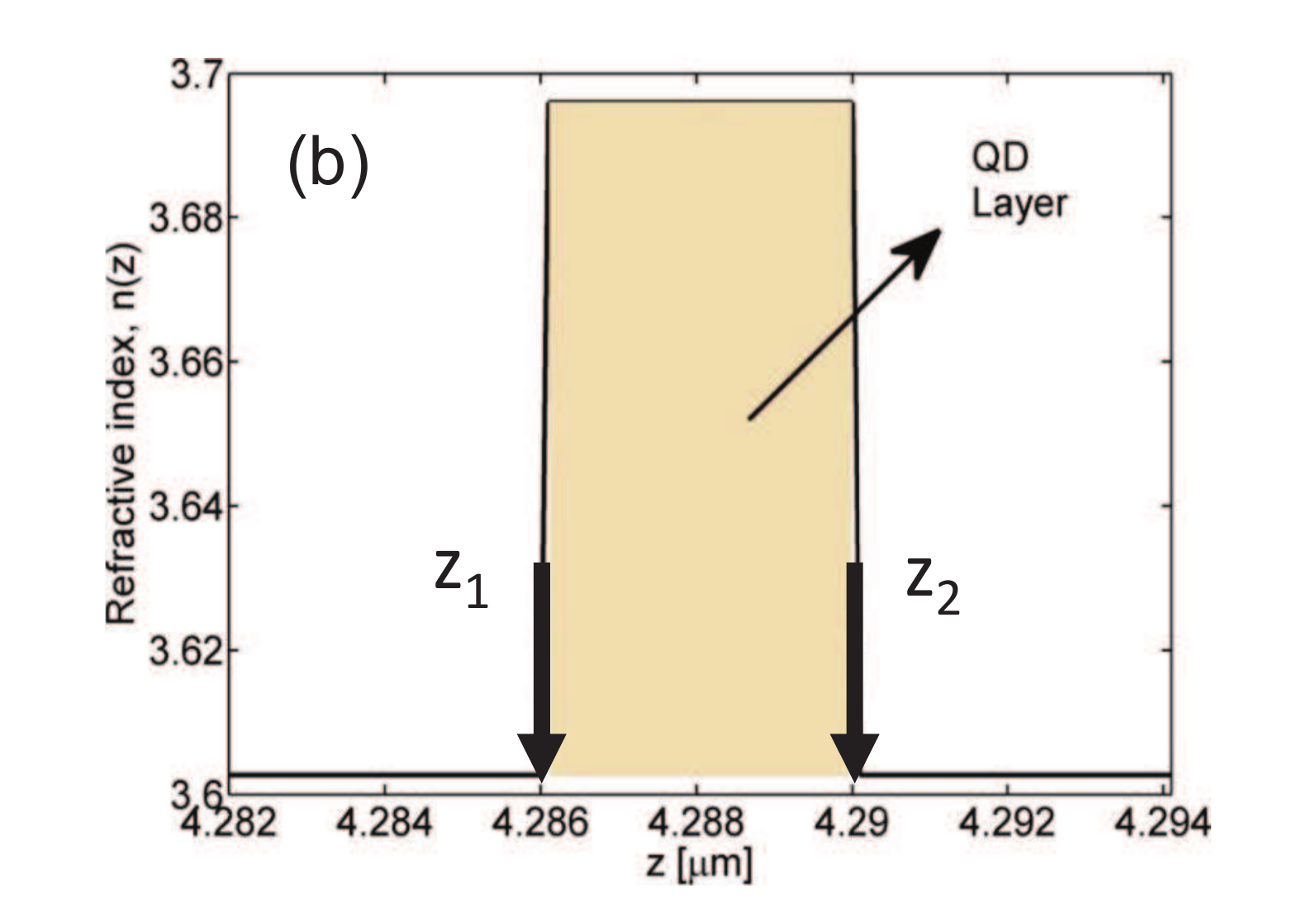}
\quad
\includegraphics[height=3 cm]{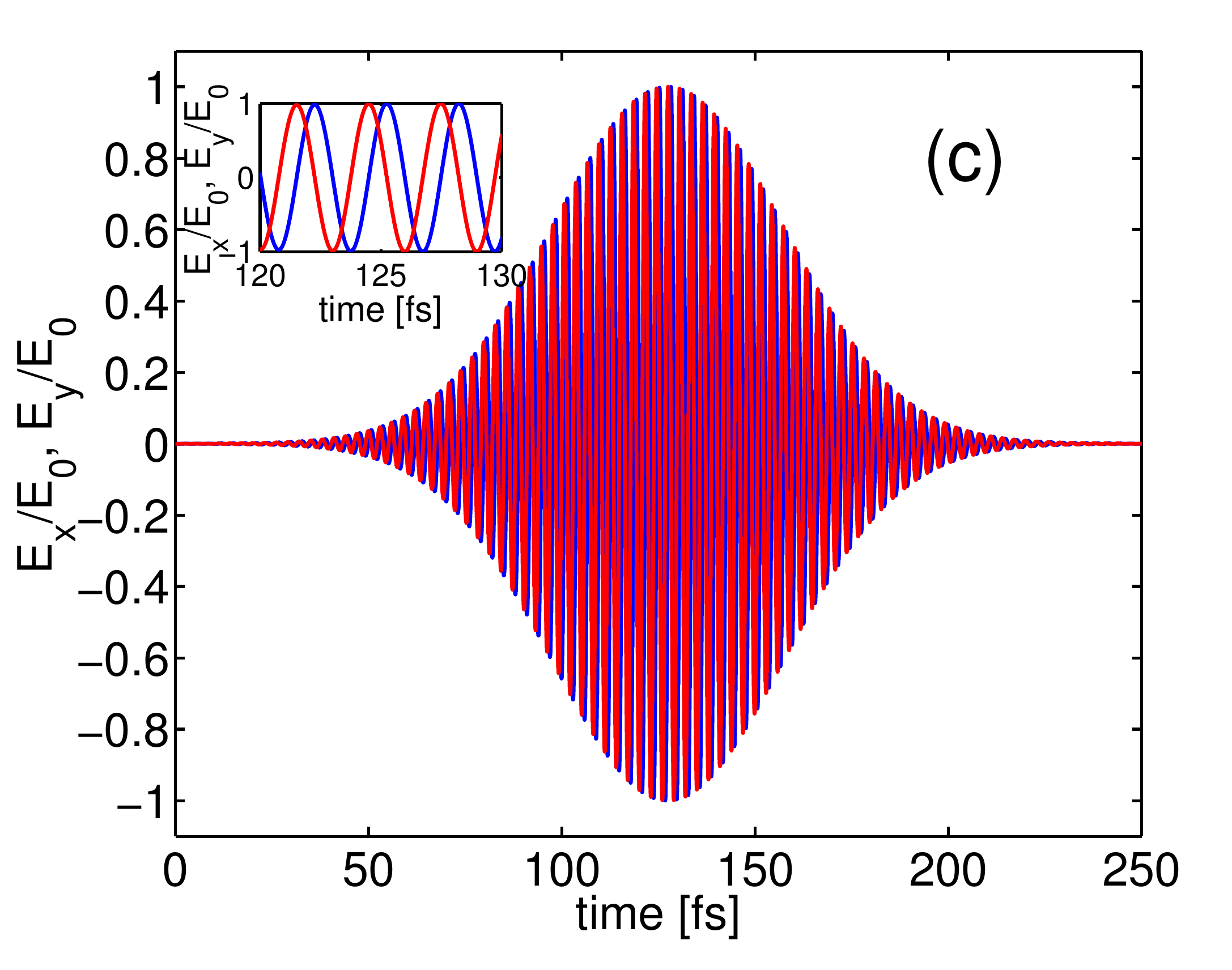}
}
\caption{(a) Refractive index profile of a micropillar structure with bottom/top DBRs consisting of $18.5/5$ $AlAs/GaAs$ pairs and a $3\lambda$-cavity (GaAs) with a modulation doped InGaAs QD layer embedded in the middle of the cavity; the micropillar is pumped from the top mirror through an optical fibre (rightmost layer with refractive index, $n_f= 1.436$); $L$ - total length of the simulation domain;(b) Zoom-in of the region around the QD layer in (a); $z_1$ and $z_2$ denote the left and right end points of the QD layer;(c) Time evolution at $t=0.334 \,\,\mathrm{ps}$ of a Gaussian left circularly polarised input pulse at the input facet $z=L$; the time dependence of the normalised $E_x$ (blue) and $E_y$ (red) field components is shown. The initial phase difference between the two $E$-field components is $\pi/2$; inset -- zoom-in of the electric field components time dependence.
}
\label{fig:geometry}
\end{figure}

The micropillar distributed Bragg reflector (DBR) geometry is implemented through spatially dependent 1D refractive index profile along the structure (see Fig.~\ref{fig:geometry} (a)). As in reality light can freely propagate and escape the cavity from the input and output cavity facets, perfectly absorbing (transmitting) boundary conditions are imposed at both $z=L$ and $z=0$. The absorbing boundary conditions are based on Engquist-Majda one-way wave equations discretised with a second-order accuracy Mur finite-difference scheme \cite{SlavchevaPRA2002}. The latter allows us to compute the cavity loss and the Q-factor of a realistic micropillar cavity, usually assumed as an external parameter in the models. In particular, we consider the micropillar cavity, consisting of $18.5/5$ pairs of alternating $AlAs/GaAs$ layers and a $3 \lambda$-cavity with an embedded modulation doped $InGaAs$ QD layer (Fig.~\ref{fig:geometry}).

The master pseudospin equations (\ref{eq:pseudospin_master}) and the Maxwell's curl equations (\ref{eq:Maxwell_vector_circ_polar}) are solved self-consistently in the time domain by employing the Finite-Difference Time-Domain (FDTD) method for a resonant Gaussian circularly polarised pulse, with carrier frequency $\omega_0$ (Fig.~\ref{fig:dot_micropillar}(b)), centred at $t_0$ with a standard deviation $t_d$, propagating and interacting with the trion transition of the QD. We consider Goursat initial boundary value problem which requires the knowledge
of the whole time history of the initial field along some characteristic, e.g., at $z=L$ at the right (top) boundary of our simulation
domain where the pulse is injected from. The pulse time dependence at $z=L$ given by:
\begin{equation}
\sigma ^ \pm  \left\{ \begin{array}{l}
 E_x \left( {z = L,t} \right) = E_0 exp\left[ { - \left( {t - t_0 } \right)^2 /t_{d}^2 } \right] \cos (\omega _o t) \\
 E_y \left( {z = L,t} \right) =  \pm E_0 exp\left[ { - \left( {t - t_0 } \right)^2 /t_{d}^2 } \right]\sin (\omega _o t) \\
 \end{array} \right. \,\,\
 \label{eq:circularly_polarised_pulse}
\end{equation}
where $E_0$ is the initial pulse amplitude and the sign $\pm$ corresponds to right(left) circularly polarised excitation. In the case of a linearly polarised excitation, we will assume an $x-$ polarised pulse, given by:
\begin{eqnarray}
\begin{array}{l}
\label{eq:linear} X \left\{ \begin{array}{l}
 E_x \left( {z = L,t} \right) = E_0 e^{ - {{\left( {t - t_0 } \right)^2 } \mathord{\left/
 {\vphantom {{\left( {t - t_0 } \right)^2 } {t_d^2 }}} \right.
 \kern-\nulldelimiterspace} {t_d^2 }}}\cos (\omega _0 t) \\
 E_y \left( {z = L,t} \right) =  0 \\
 \end{array} \right.
\end{array}
 \label{eq:lin_polarised_pulse}
\end{eqnarray}

\section{Numerical results}
\label{sec3}
In the following, we define the model parameters. We take the trion transition energy, $\hbar \omega_0=1.388\,\,\mathrm{eV}$, corresponding to a resonant wavelength, $\lambda_0= 894\,\,\mathrm{nm}$ and the trion spontaneous emission time of $1/\Gamma \sim 820\,\,\mathrm{ps}$ from \cite{Ruth2016}. Using these parameters, the optical transition dipole matrix element is calculated, giving $\wp \sim 0.57 \mathrm{\,e \cdot nm}$. The longitudinal electron spin-flip relaxation time $1/\gamma_1 \sim 500 \,\,\mathrm{ps}$ (see lower blue/grey curved arrows in Fig.~\ref{fig:dot_micropillar}(b)) is calculated in \cite{Merkulov2002}$^{,}$\cite{Khaetskii2002}. The hole spin-flip relaxation time in the trion state (see upper blue/grey curved arrows in Fig.~\ref{fig:dot_micropillar}(b)) is due to phonon-assisted processes and has been measured on the order of $1/\gamma_3 \sim 170 \,\,\mathrm{ps}$ \cite{Greilich}. An estimate for the electron spin decoherence time, $1/\gamma_2 \sim 450 \,\,\mathrm{ps}$, is obtained from the experimentally measured value \cite{Economou2005} and the trion-state hole spin decoherence time is assumed on the order of $1/\Gamma _{\tau} \approx 1/{2\gamma _3}= 340 \,\,\mathrm{ps}$.

In order to model a single dot, we make use of the ergodic hypothesis which states that the time average of an observable is equivalent to an ensemble average over a large number of replicas of the quantum system, thereby allowing to predict single quantum system properties on the basis of macroscopic averages of observables. We consider an ensemble of charged QDs with resonant dipole density, $N_d$, and select $N_d \sim 3.18 \times 10^{24}\,\,\mathrm{m^{-3}}$, to give on average one dot within the QD volume (for a typical dot with a diameter $d=10 \,\,\mathrm{nm}$ and height $h=4 \,\,\mathrm{nm})$, thereby restricting the simulation to a microscopic volume containing a single dot.

We choose an ultrashort pulse with pulse duration $T_p=100 \,\,\mathrm{fs}$ and an area of $\pi$ which completely excites the spin population of the initial state into the excited trion state (Fig.~\ref{fig:geometry}(c)). From these parameters, using the value for the dipole matrix element $\wp$ above, we can calculate the initial pulse amplitude of a Gaussian pulse, giving $E_0=1.44 \times 10^8 \,\,\mathrm{V \cdot m^{-1}}$. The corresponding Rabi frequency is then $\Omega _R  = \frac{{\wp E_0 }}{\hbar }\sim 1.245 \times 10^{14} \,\,\mathrm{rad/s}$, giving a coupling rate $g \approx 1.98 \times 10^{13}\,\,\mathrm{s^{-1}}  >> \Gamma, \Gamma_{\tau},\gamma_i, \,\,\ i=1,2,3$ all of which vary in the interval $(2 \div 6) \times 10^{9} \,\,\mathrm{s^{-1}}$. Although this emitter-photon coupling rate is larger than the spin relaxation and decoherence rates, as we shall show below, the micropillar cavity is operating in the weak-coupling regime due to cavity losses exceeding the coupling rate. In addition, the trion resonance is spectrally detuned with respect to the cavity mode. In what follows, we study three different scenarios of resonant polarised excitation of the coupled dot-cavity in the pulsed regime and compute the quantum evolution of the system.

\subsection{Quantum evolution upon circularly polarised excitation}
\label{sssec:3.1}

\begin{figure}
\resizebox{.9\textwidth}{!}{%
\includegraphics[height=5 cm]{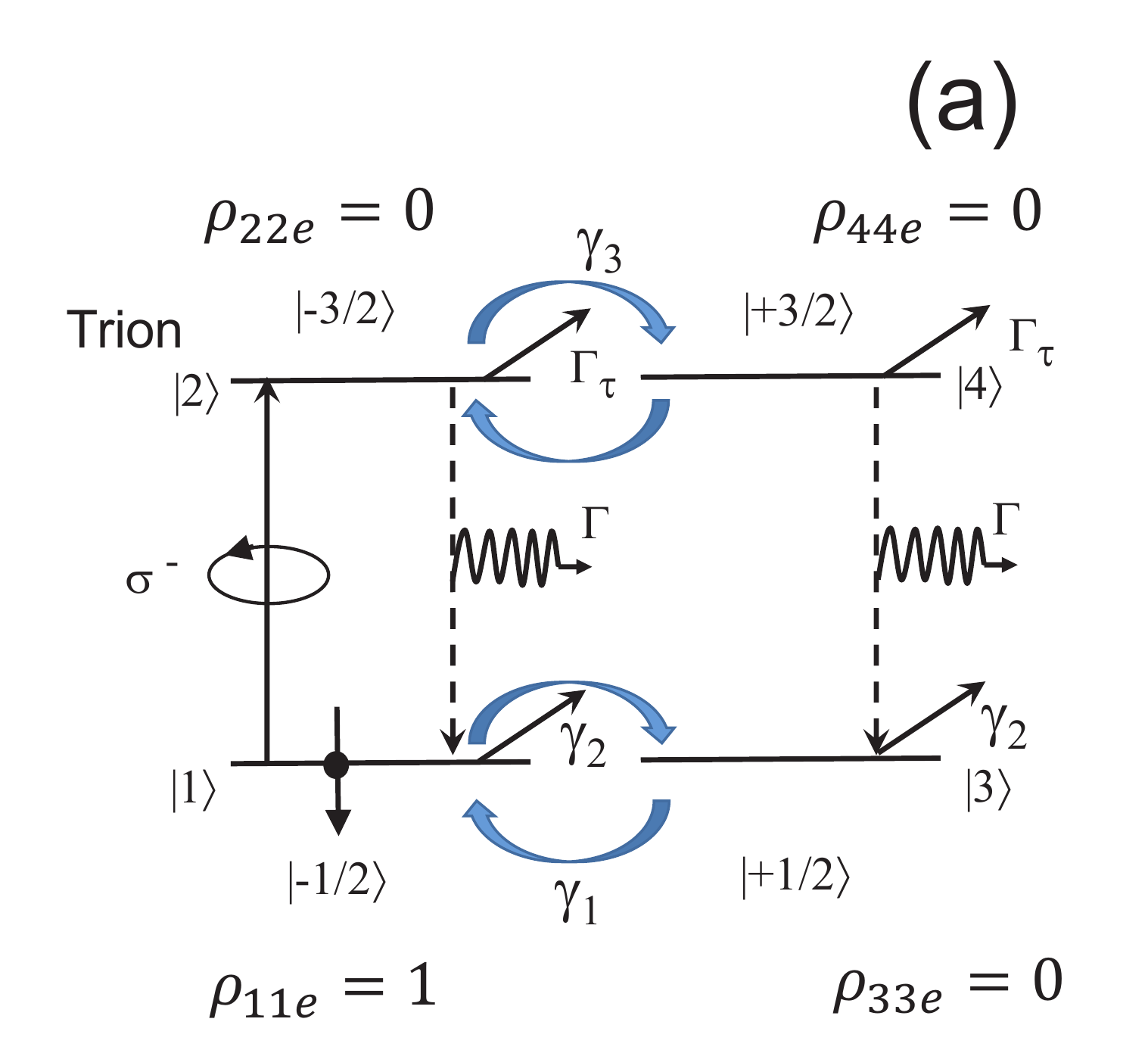}
\quad
\includegraphics[height=5 cm]{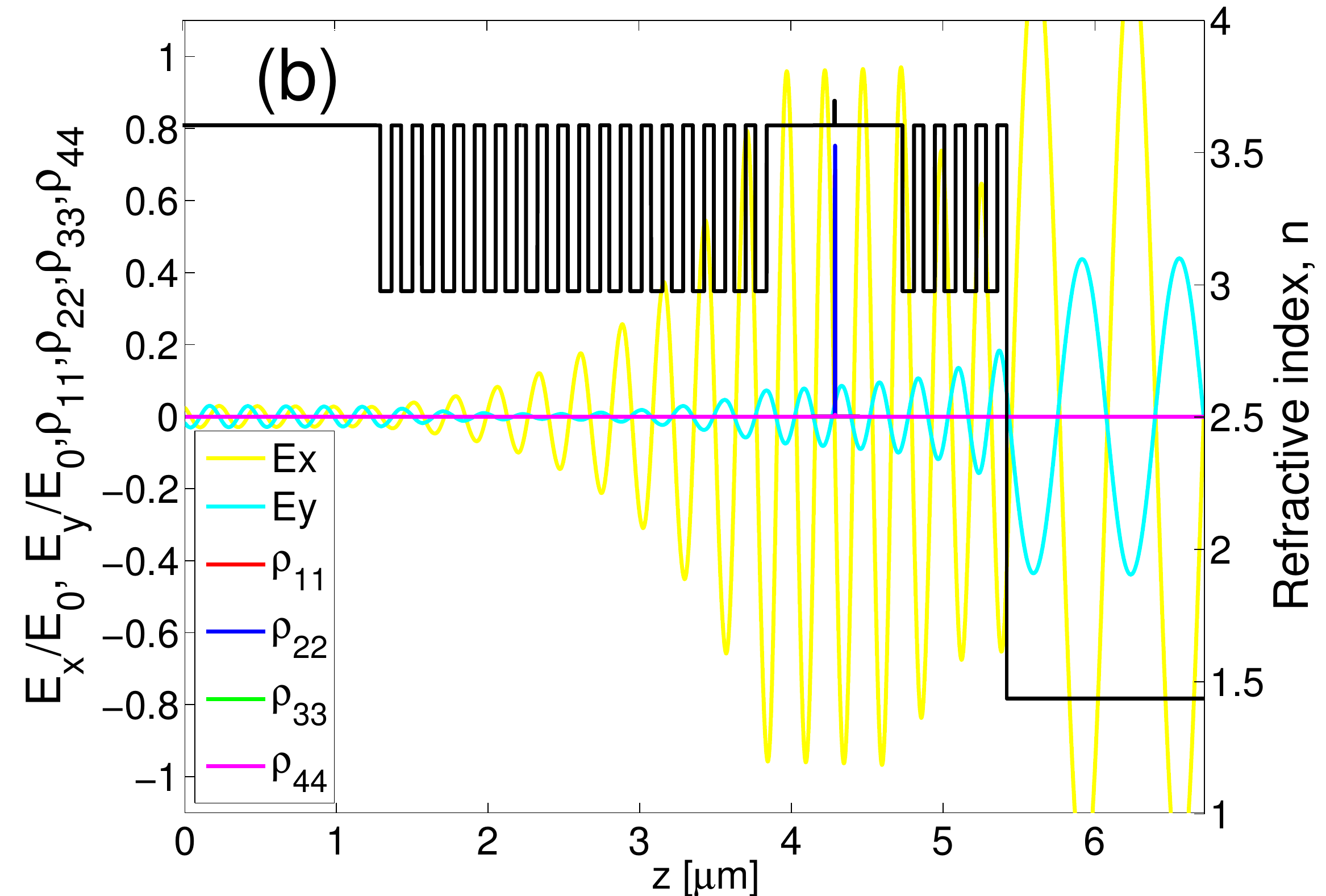}
\quad
\includegraphics[height=5 cm]{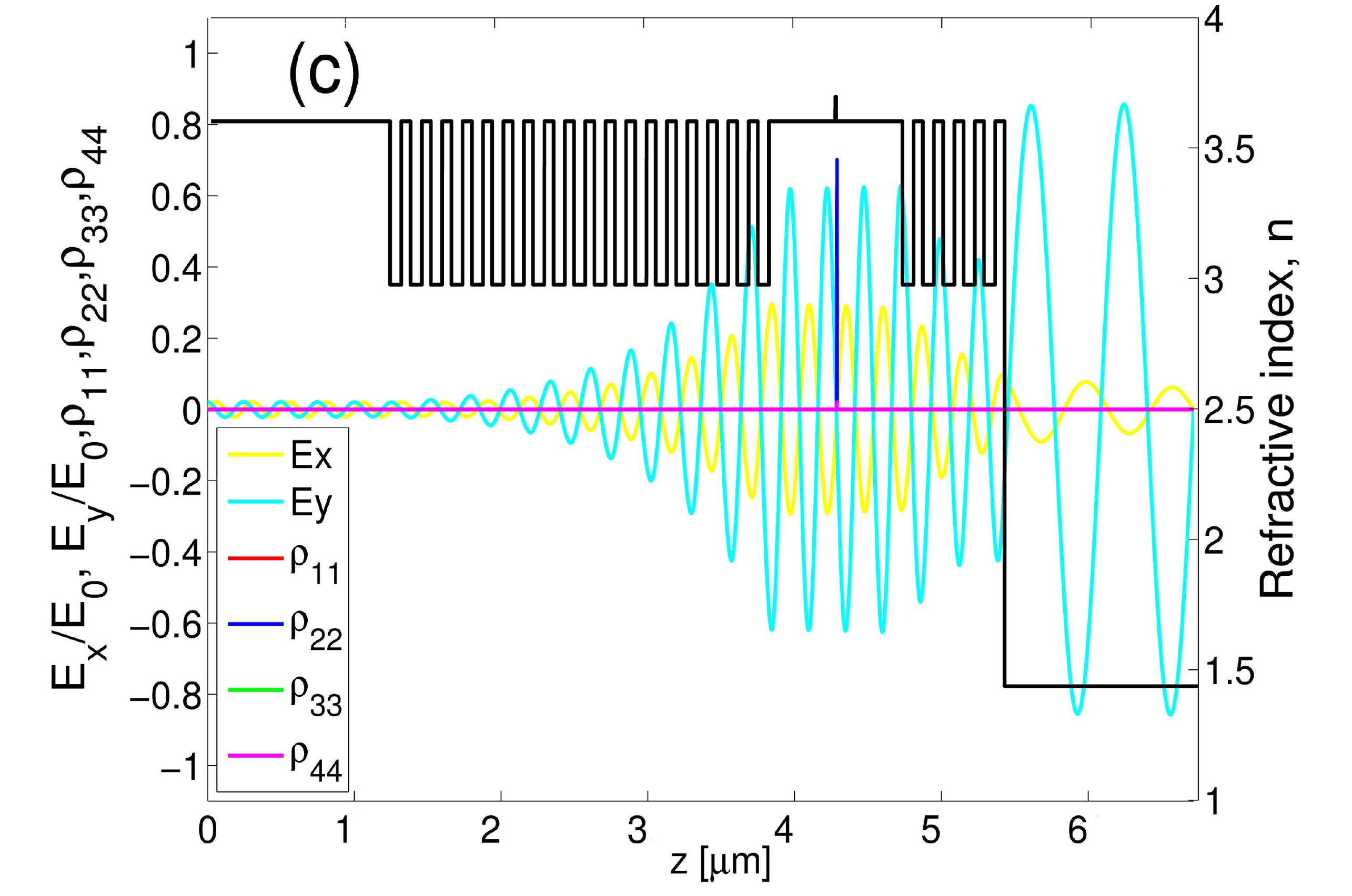}
}
\caption{(a) Excitation scheme under left-circularly  ($\sigma^{-}$) polarised pulse; initial boundary conditions: entire spin population prepared in spin-down state $|1\rangle$ by e.g. optical pumping; $\rho_{iie}\,\,\ i=1,...,4$ are the equilibrium level populations; Refractive-index profile (black line) of the micropillar cavity with embedded QD, and a snapshot at $t=3.33 \,\,\mathrm{ps}$ (b) of the $E$-field components (yellow and cyan curves) and level populations (red, blue, green and magenta) superimposed on it, (c) snapshot at $t=13.34 \,\,\mathrm{ps}$ showing amplitude build-up of both circularly polarised pulse components within the cavity.}
\label{fig:spatial_dynamics}
\end{figure}

Consider initial spin-down population prepared entirely in ground level $|1\rangle$ e.g. by optical spin orientation by a circularly polarised pump. We shall be interested in the quantum evolution of the four-level system described by Eqs. (\ref{eq:pseudospin_master})-(\ref{eq:polarisation}) when the trion transition is resonantly driven by a left-circularly polarised pulse, $\sigma^{-}$ (see excitation scheme in Fig.~\ref{fig:spatial_dynamics} (a)). A snapshot of the spatial dynamics of the $E$-field components and the level populations at the time $t=3.33 \,\,\mathrm{ps}$ is shown in Fig.~\ref{fig:spatial_dynamics} (b). The amplitude of the $E_x$ circularly polarised field component at this particular time moment is already accumulated within the cavity exhibiting a standing-wave profile, while the $E_y$ component is still decaying within the cavity. This is due to the phase shift of $\pi/2$ between the two orthogonal components (note the phase shift between the two components - when $E_x$ has a maximum, $E_y$ has a minimum and vice versa).The spin population of level $|2\rangle$, $\rho_{22}$ (blue line) is excited by the incident pulse and the population transfer between all four levels is initiated. At later times, the $E_x$ and $E_y$ amplitudes gradually increase and the fields become localised within the cavity, thereby exhibiting standing-wave profiles (Fig.~\ref{fig:spatial_dynamics} (c)). At much later times, the amplitude of both fields decreases due to cavity photon loss through the DBRs (not shown), however spin population transfer between the levels still occurs due to the incoherent relaxation processes acting on longer timescales. The spatial dynamics shows the interplay between the interference cavity effects and the resonant absorption or amplification of the $E$-field components: the absorbed or amplified by the QD trion transition light is emitted back to the cavity and thus alters the interference pattern. The model thus accounts for the feedback effects within the cavity.

\begin{figure}
 \resizebox{.9\textwidth}{!}{%
\includegraphics[height=4cm]{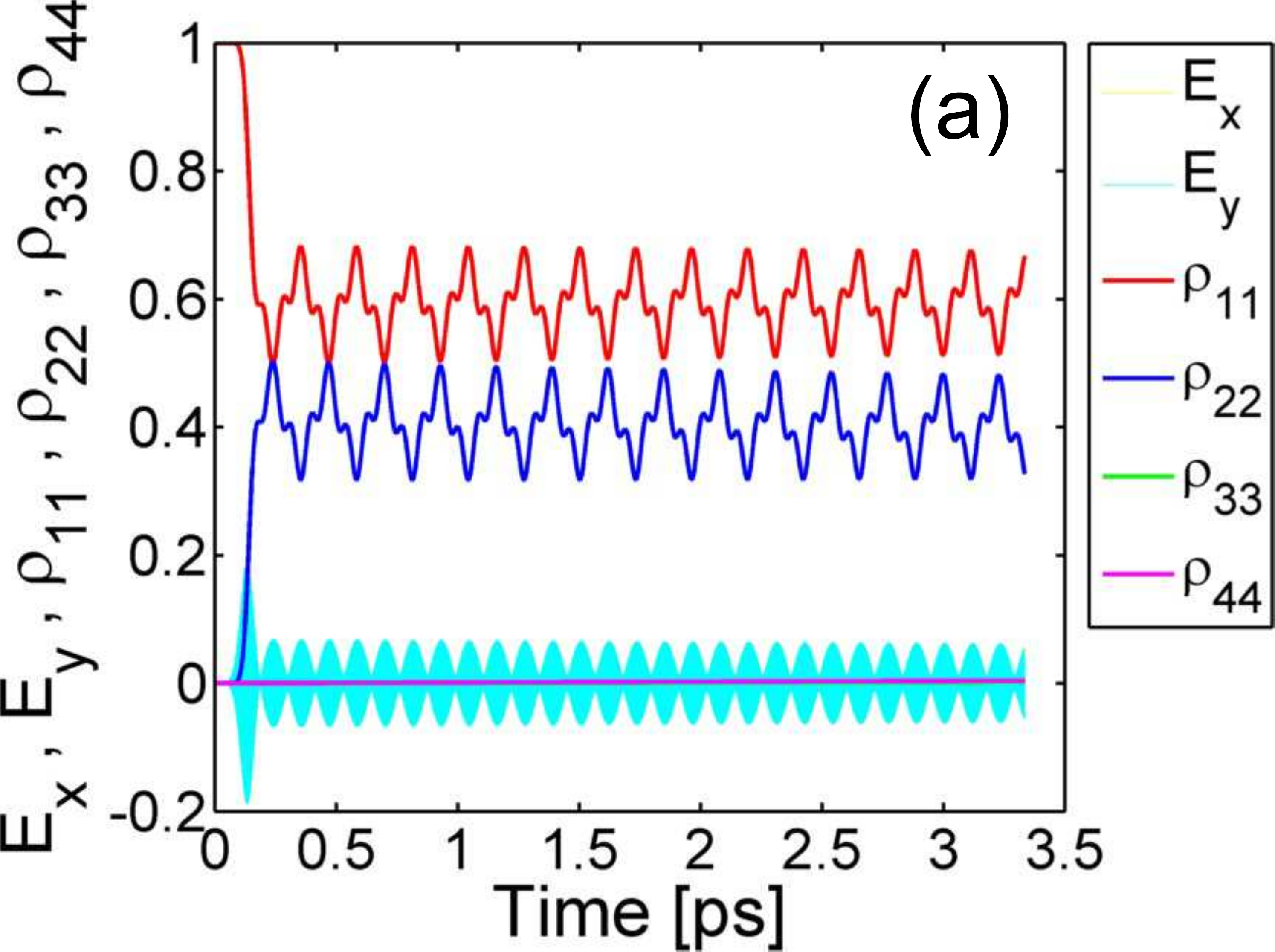}
\includegraphics[height=4cm]{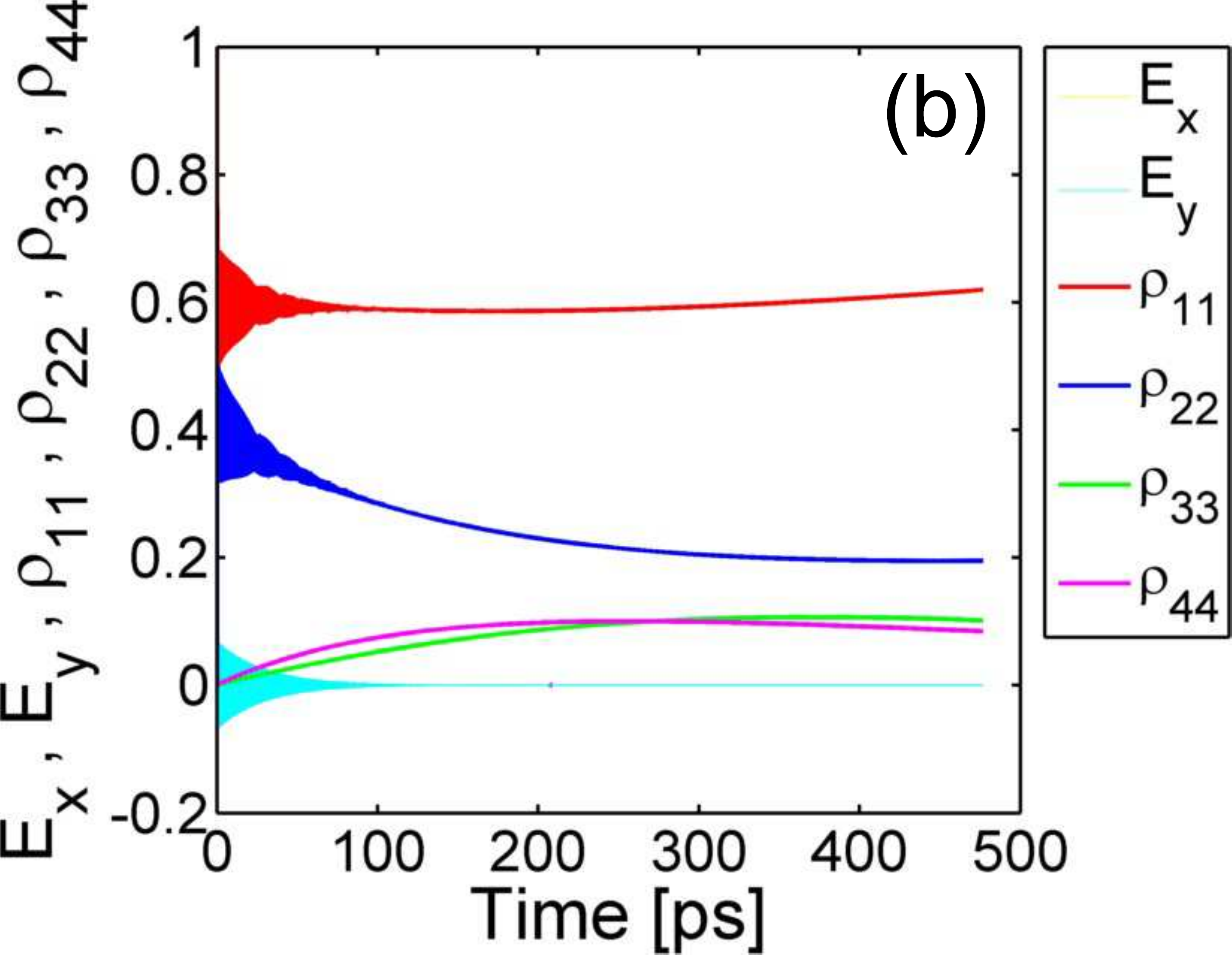}
}
\resizebox{.9\textwidth}{!}{%
\includegraphics[height=4cm]{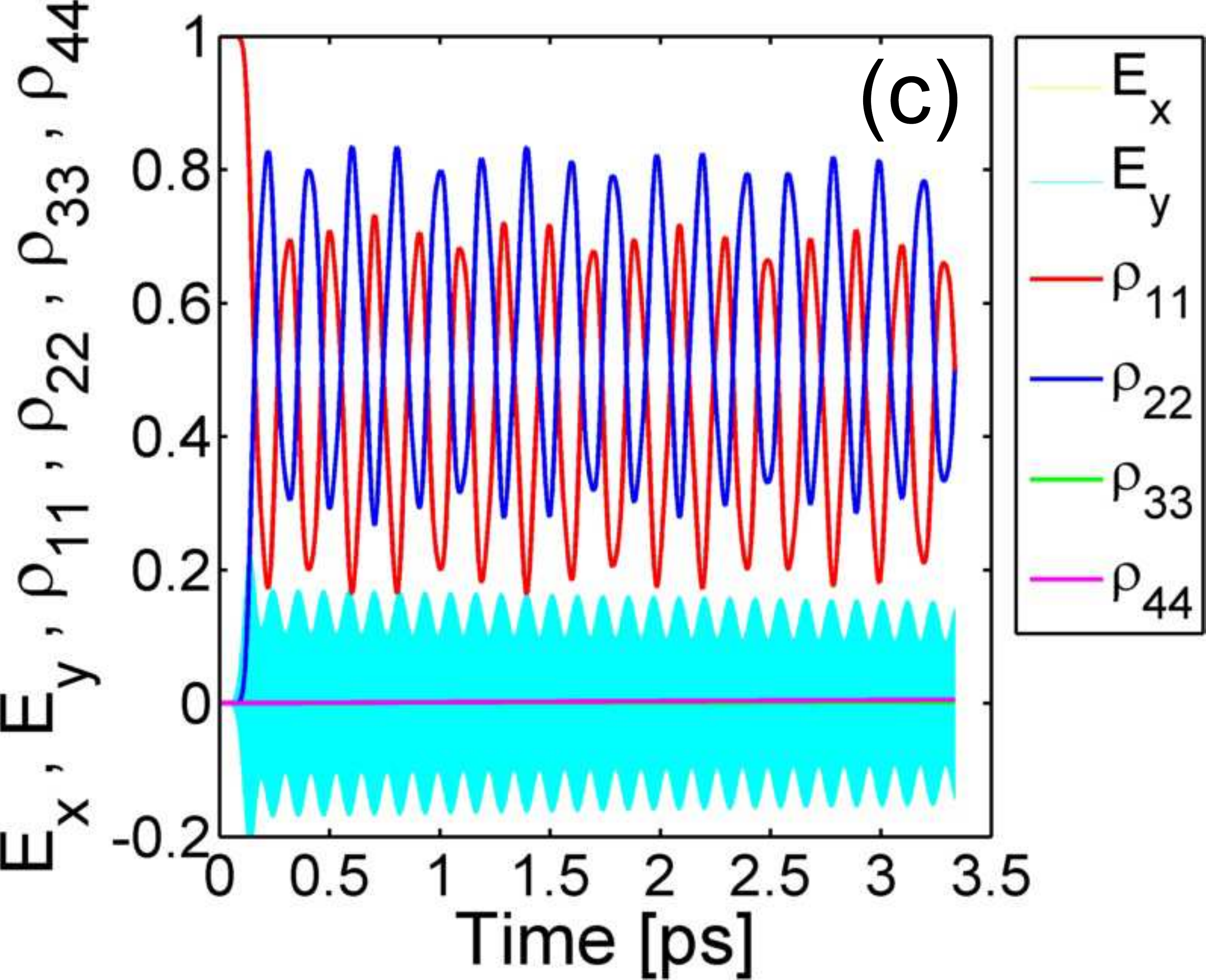}
\includegraphics[height=4cm]{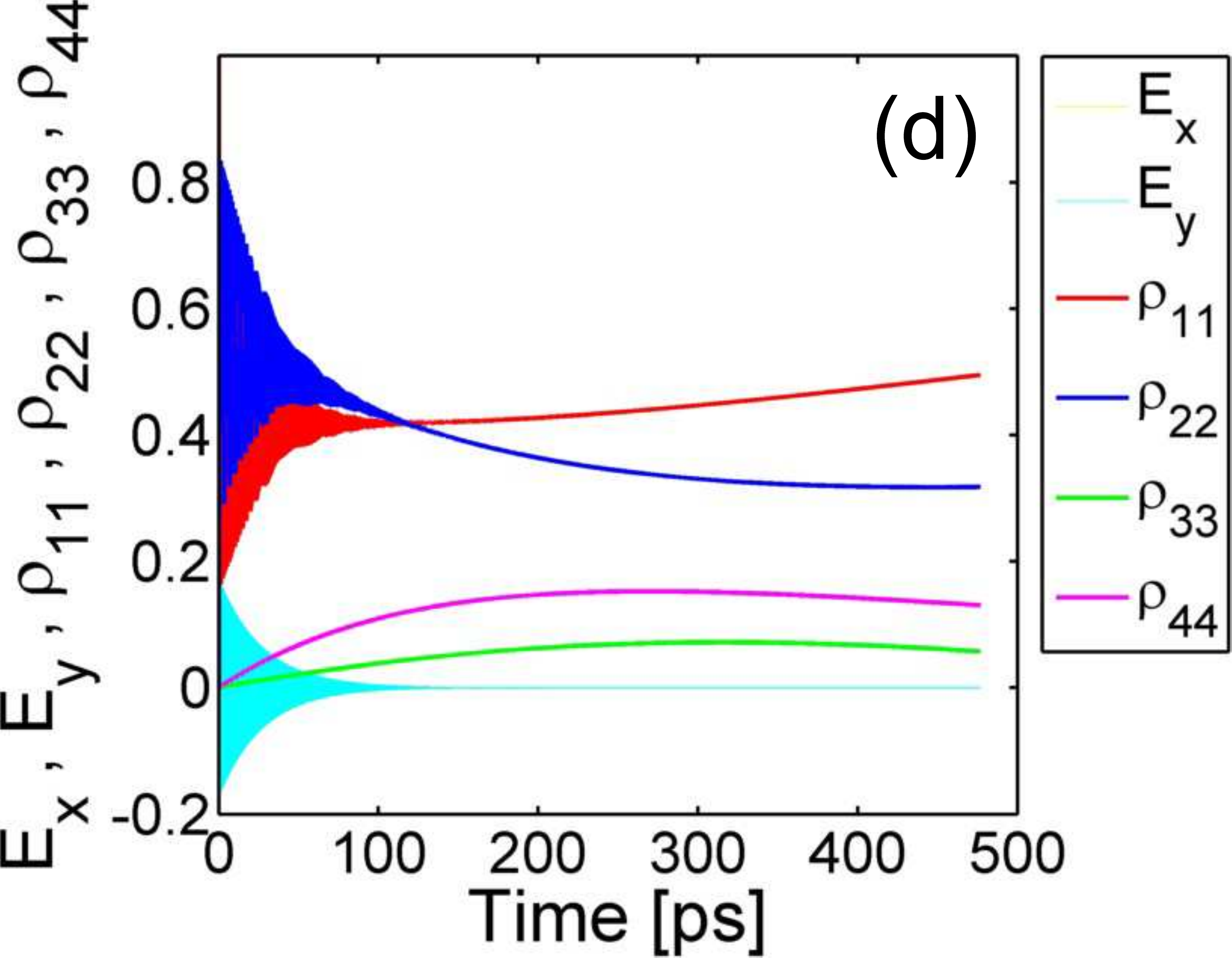}
}
\caption{Time evolution of the normalised (with respect to the initial pulse amplitude $E_0$) electric-field components of a $\sigma^{-}$ pulse and level populations at the left/right end points of the QD layer (see Fig.~\ref{fig:geometry}(b)) for a spin-down initial state:
(a) ultrashort dynamics; (b) long-time dynamics at $z_1$; (c) ultrashort dynamics; (d) long-time dynamics at $z_2$. The population of level $|2\rangle$ (blue curves) is proportional to the experimentally detected time-resolved photoluminescence at the trion transition. $E_x$ component (yellow) is not visible as it is overlapped by the $E_y$ component (cyan) with the same amplitude.}
\label{fig:time_dynamics}
\end{figure}

The temporal dynamics is calculated at the end points, $z_1$ and $z_2$, of the QD layer displayed in Fig.~\ref{fig:geometry} (b). The short-time (up to $3.5\,\,\mathrm{ps}$) and long-time time (up to $500 \,\,\mathrm{ps}$) traces of the electric field components and the populations of all four levels are displayed in Fig.~\ref{fig:time_dynamics} sampled at the beginning, $z_1$ (a),(b) and the end, $z_2$ (c),(d) of the QD layer. Initially, the $\pi$-pulse excites the population residing in level $|1\rangle$ to the excited level $|2\rangle$. The polarised time-resolved photoluminescence detected on the trion transition is proportional to the population, $\rho_{22}$ of level $|2\rangle$ (blue curve in Fig.~\ref{fig:time_dynamics}). Note that although $\rho_{22}$ looks as saturated at $t \approx 500 \, \mathrm{ps}$, similar to the bare QD \cite{SlavchevaPRB2008}, it still continues to decay very slowly (on a much longer time scale of a few $\mathrm{ns})$ to its equilibrium value of $0$, far beyond the trion recombination time of $820 \,\mathrm{ps}$. The level populations $\rho_{11}$ and $\rho_{22}$  exhibit fast beatings due to incomplete damped population Rabi flopping between level $|1\rangle$ and $|2\rangle$ (Fig.~\ref{fig:time_dynamics}(a,c)).

\begin{figure}
\resizebox{.9\textwidth}{!}{%
\includegraphics[height=5 cm]{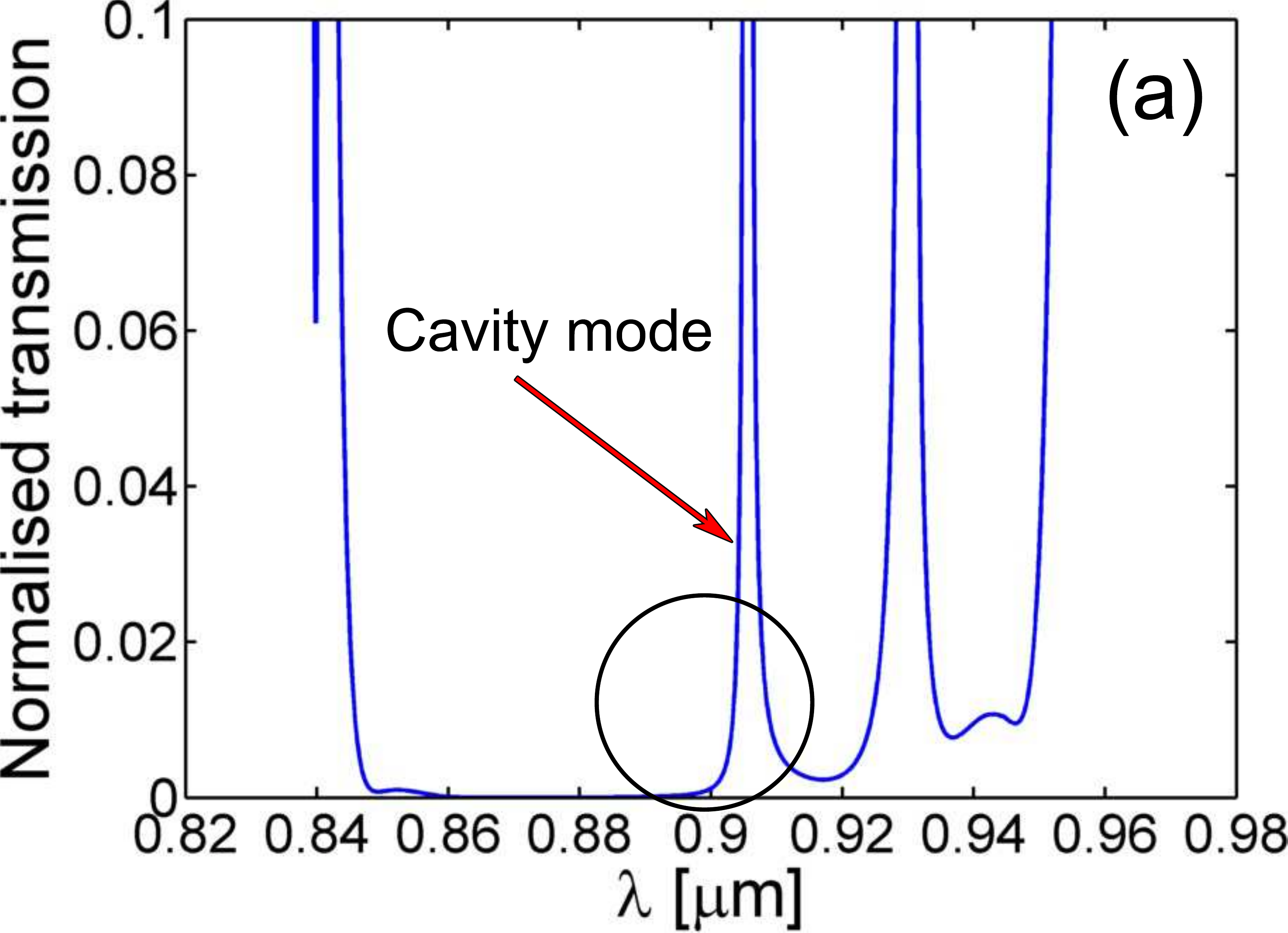}
\includegraphics[height=5 cm]{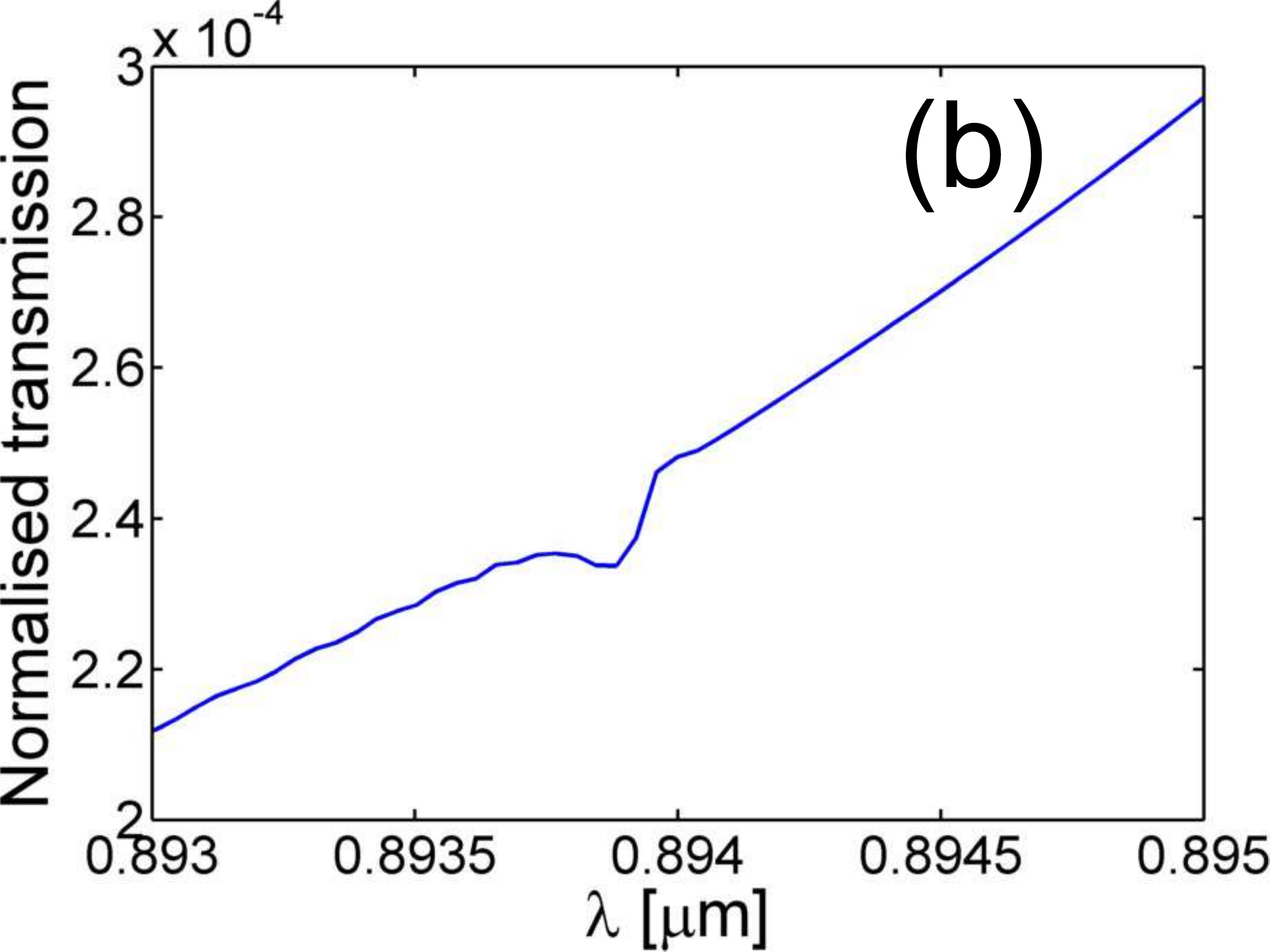}
\includegraphics[height=5 cm]{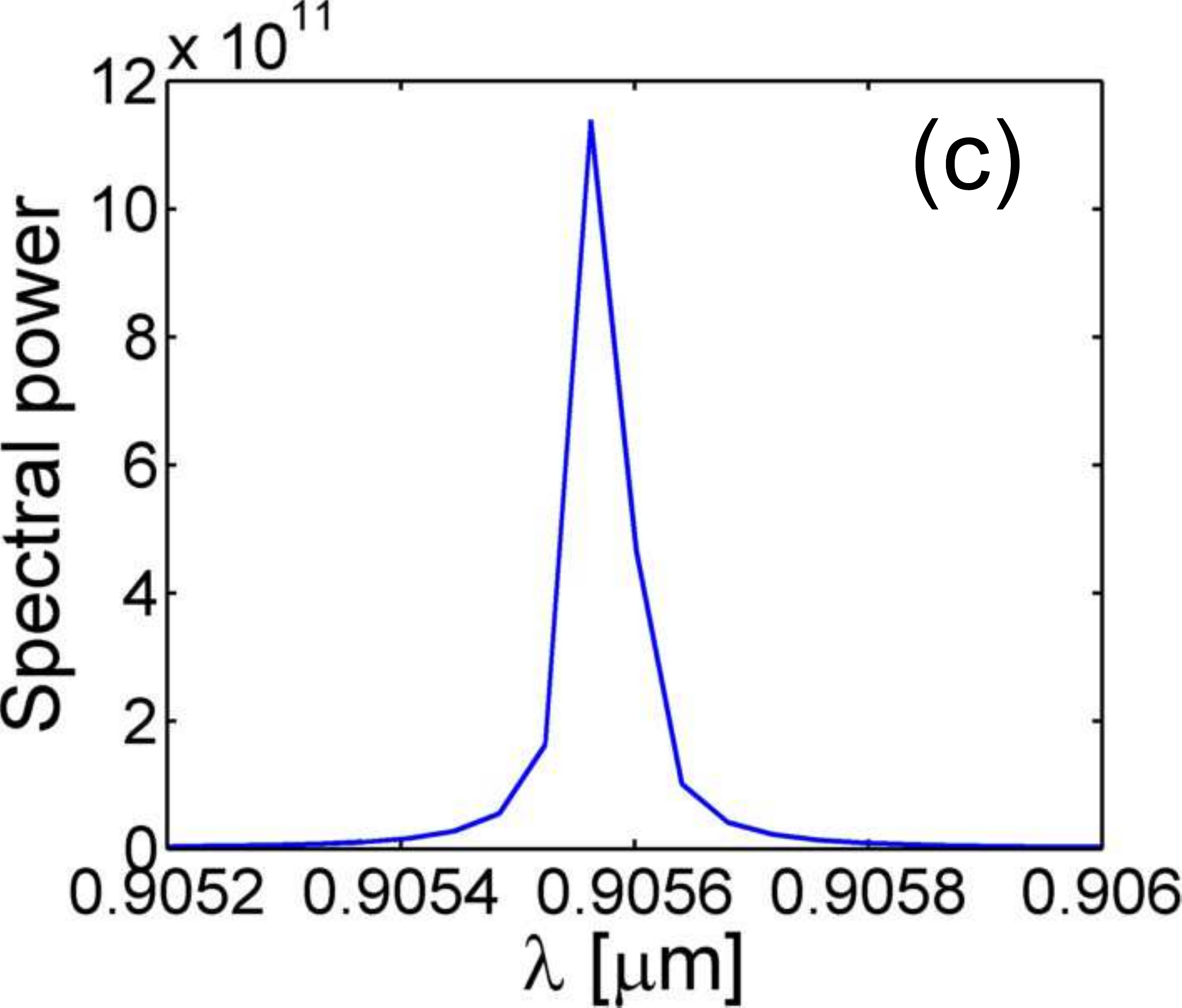}
}
\caption{(a) Transmission spectrum of the micropillar cavity as a function of the photon wavelength under left-circularly polarised $\sigma^{-}$  pulse excitation showing the cavity stop-band and the fundamental cavity mode at $\lambda=0.9055 \,\,\mathrm{\mu m}$; (b) Zoom-in of the region around the trion resonance denoted ba a circle in (a), showing a resonant absorption dip at the QD trion transition wavelength $\lambda=0.894 \,\,\mathrm{\mu m}$; (c) Fundamental cavity mode; the cavity loss is evaluated from the FWHM of the transmission peak.}
\label{fig:transmission_spectrum}
\end{figure}

The FDTD approach allows us to compute the optical transmission spectrum of the active micropillar structure under an ultrashort (broadband) pulse excitation. By taking the Fourier transform of the time trace sampled at the structure output facet (at $z=0$) and normalising it with respect to the initial pulse spectrum, one can obtain the cavity stop band and the cavity modes, shown in Fig.~\ref{fig:transmission_spectrum}(a). The micropillar structure is asymmetric (the number of DBR layers in the bottom and top mirrors largely differ) and cavity mode is detuned from the QD transition, confirming that the cavity operates in the weak coupling regime.

A zoom-in of the fundamental cavity mode in the vicinity of the QD trion resonance is displayed in Fig.~\ref{fig:transmission_spectrum}(b). Similar to the experimentally observed reflectivity peak in \cite{ArnoldNatComm2015}, the dip in the calculated transmission spectrum is a signature of resonant absorption of the circularly polarised pulse by the QD trion transition. The resonantly excited trion ground transition generates a dipole field that interferes coherently with the exciting pulse field. We have previously demonstrated numerically resonant absorption (gain) in a two-level system which manifests itself as a dip (peak) superimposed on the broadband pulse spectrum, depending on whether the system is initially prepared in its ground (excited) state \cite{Slavcheva_PRA2005}. In the dot-micropillar case, the transmission dip is superimposed on the cavity mode. The cavity loss of the realistic micropillar cavity can be calculated from the FWHM of the cavity mode peak displayed in Fig.~\ref{fig:transmission_spectrum}(c), giving $1/\kappa=1.25 \,\,\mathrm{ps}$. Therefore, $\Omega _R  = \frac{{\mu E_0 }}{\hbar } \ll \kappa$ and we conclude that the cavity is operating in the weak coupling regime.

\subsection{Phase shift calculation}
\label{sssec:3.2}

In order to calculate the resonant pulse optical rotation angle, we make use of the complex propagation factor,$e^{i \vec k_c . \vec l}$,  of the electromagnetic wave in an absorbing/amplifying medium describing the QD layer with a thickness $l=z_2-z_1$. Here we have defined a complex propagation wave vector $k_c=\beta+i\gamma$, where $\beta$  and $\gamma$ are the phase shift and the gain/absorption coefficient over the QD layer thickness. We calculate the Fourier transform of the time trace of the $E_x$ and $E_y$ electric field components sampled at the two ends, $z_1$ and $z_2$, of the QD layer.  Then by setting $u_{x,y}  = \frac{{E_{x,y} \left( {z_2 ,\omega } \right)}}{{E_{x,y} \left( {z_1 ,\omega } \right)}} = e^{ik_{cx,y} l}$, we can define absorption (gain) coefficients and phase shifts for $E_x$ and $E_y$ field components respectively, according to:
\begin{equation}
\begin{array}{l}
 \gamma _{x,y}  =  - \frac{{\ln \left| {u_{x,y} } \right|}}{l} \\
 \Delta \varphi _{x,y}  = \beta _{x,y} l,\,\,\,\,\,\,\beta _{x,y}  = \frac{1}{l}\arctan \left( {\frac{{{\mathop{\rm Im}\nolimits} \left( {u_{x,y} } \right)}}{{{\mathop{\rm Re}\nolimits} \left( {u_{x,y} } \right)}}} \right) \\
 \end{array}
\end{equation}

\begin{figure}
\includegraphics[height=5 cm]{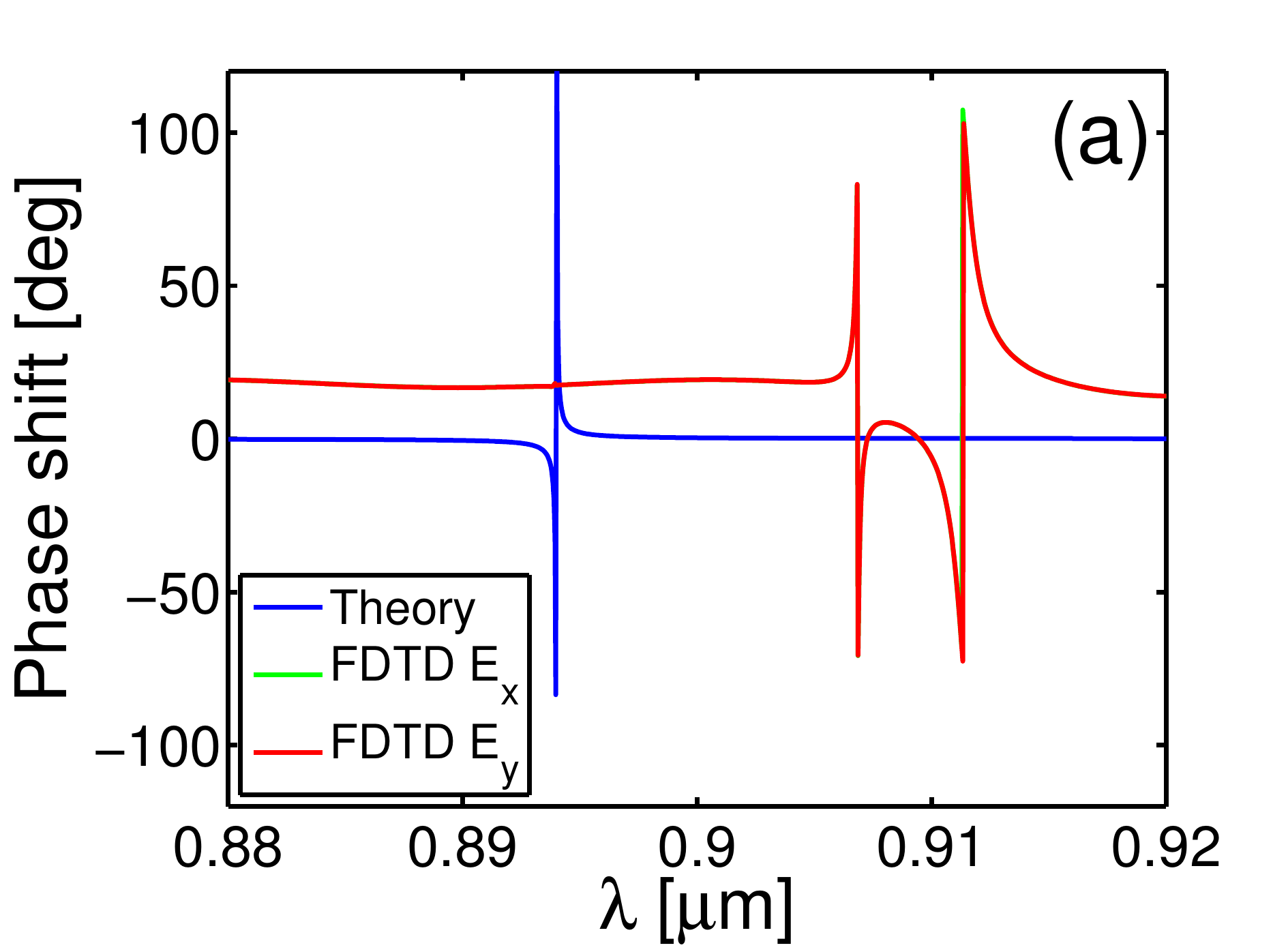}
\includegraphics[height=5 cm]{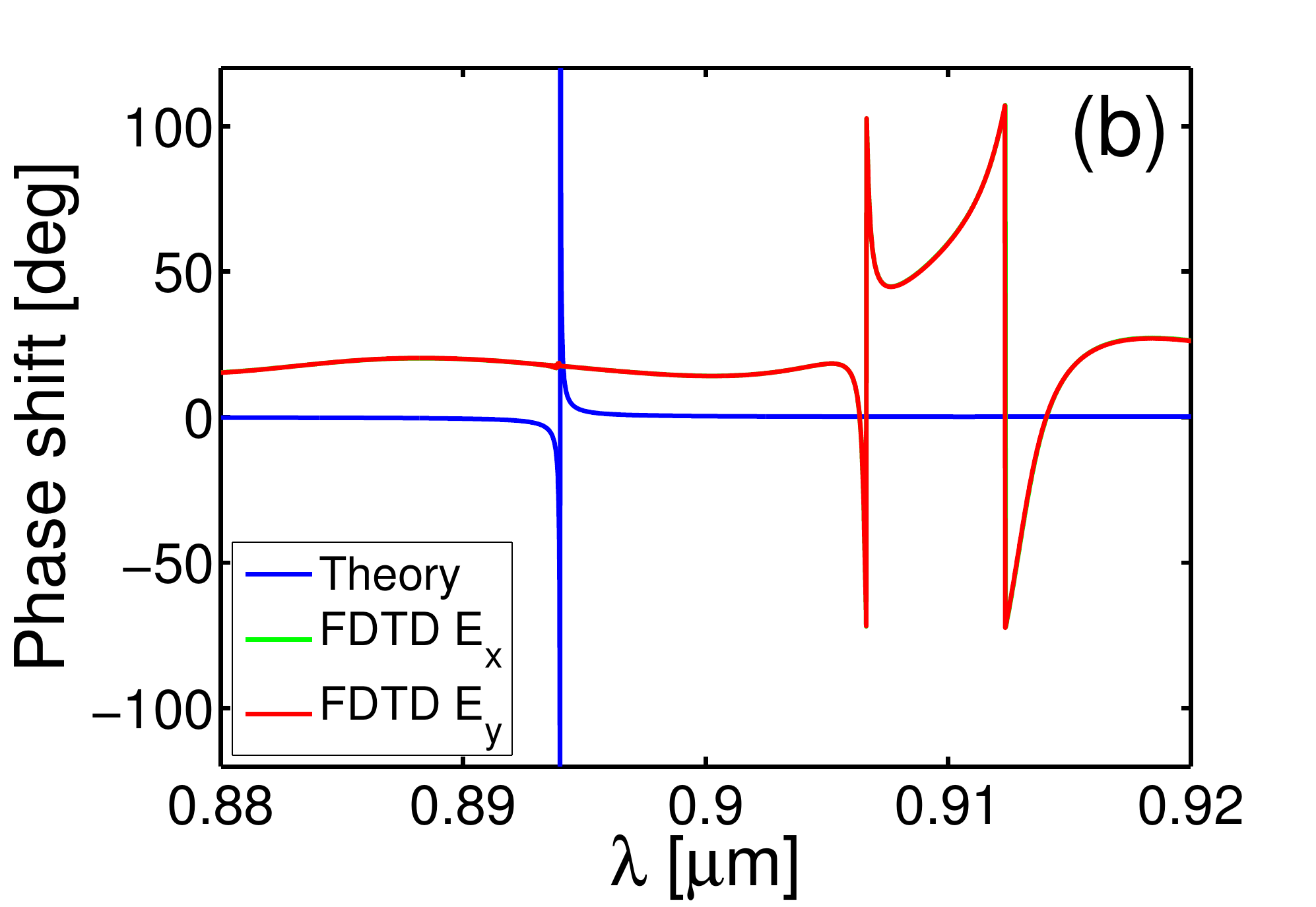}
\includegraphics[height=5 cm]{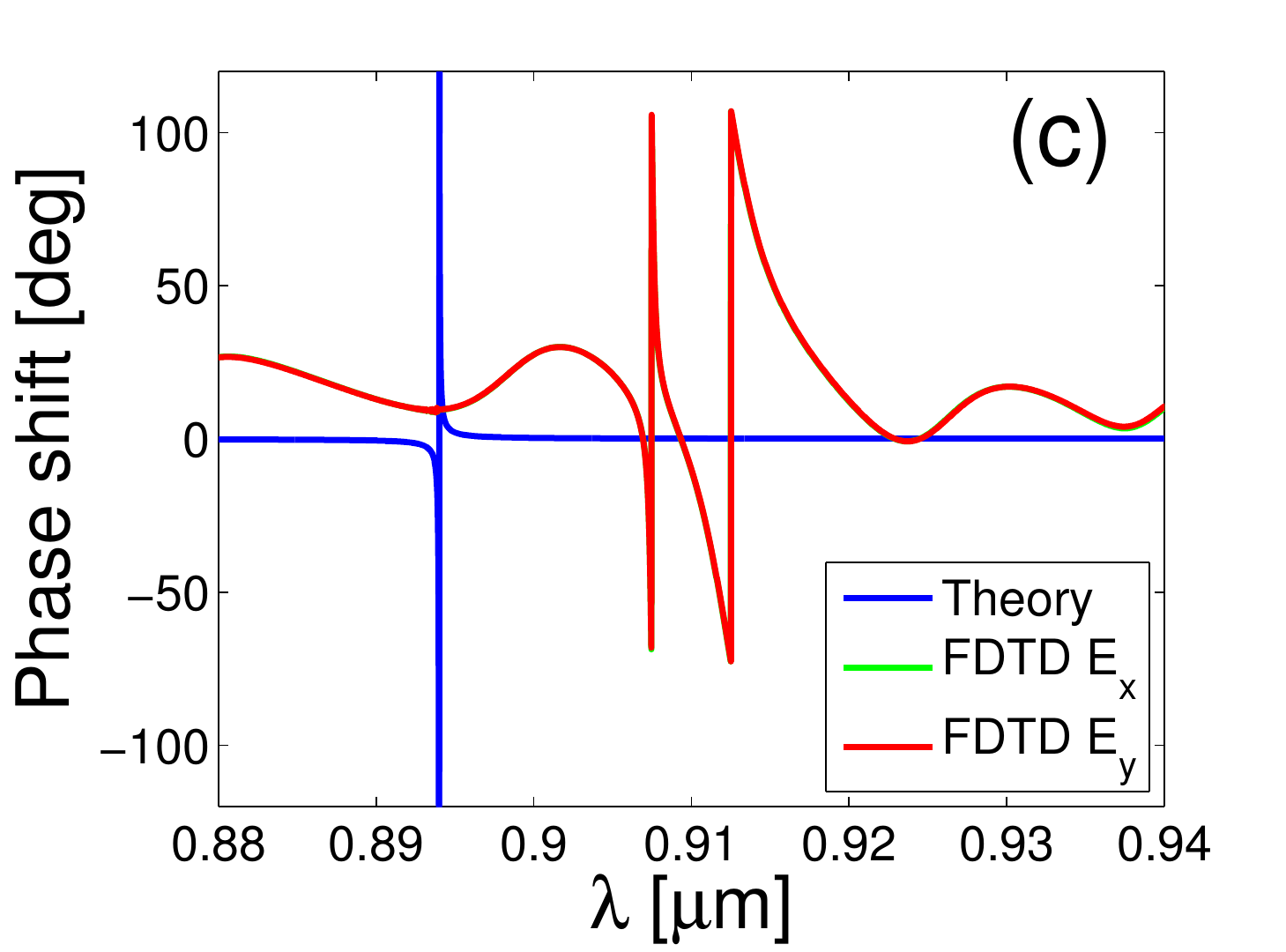}
\caption{Phase shift in degrees of the $E_x$ (green curve) and $E_y$ (read curve) pulse components of a circularly polarised $\sigma^{-}$ pulse in a micropillar cavity with bottom (top) DBRs consisting of (a) 18.5(5) periods, described in \cite{Ruth2016}; (b) $22(5)$ pairs; (c) $25(5)$ pairs. Note that the phase shifts for both $E$-field components overlap and are almost indistinguishable; Blue curve: phase shift resulting from the stationary solution of the density-matrix equations for a two-level system (no cavity) with resonant transition at $\lambda_0=894 \,\,\mathrm{nm}$; A giant polarisation rotation angle of $\pm \pi/2$ is induced by the confined spin.}
\label{fig:phase_shift_circ}
\end{figure}

For the purpose of comparison with the phase shift induced by a two-level system, we calculate the phase shift from the stationary solutions of the density matrix equations of a homogeneously broadened two-level system (see e.g. \cite{Yariv}). The results for the phase shift of an ultrashort circularly polarised pulse resonant with the trion transition in a charged QD are shown along with the phase shift induced by a bare two-level system (without a cavity) in Fig.~\ref{fig:phase_shift_circ}.

We note that two phase features occur (as opposed to the two-level system case in blue) due to coherent coupling of the pulse with the four-level system describing the QD trion resonance(cf. \cite{HuPRB2008}). The results numerically demonstrate a giant phase shift of $\pm \pi/2$, induced by the single spin confined in the QD in a realistic micropillar cavity operating in the weak coupling regime. To confirm the effect, we perform calculations on another two micropillars with a different number of DBR periods. The phase shift curve of the cavity-dot system  is red-shifted with respect to the bare QD one described by a two-level system (blue curve). This shift is $\Delta \lambda \approx \lambda_c-\lambda_0 = 13 \,\,\mathrm{nm}$ and is approximately equal to the detuning between the cavity mode and the QD trion resonance and therefore, the obtained shift in the phase shift spectra may be attributed to this detuning. Note that the cavity-dot detuning, however, is much smaller than the broadband pulse spectral width, $FWHM \approx 60 \,\,\mathrm{nm}$ and thus the pulse spectrum is encompassing both the cavity and the dot trion transition linewidths.

The detection of the Faraday rotation is usually performed in linear polarisation in pump-probe experiments, whereby the probe is linearly polarised. The linear polarisation may be decomposed into left and right circularly polarised components. Similar to Ref. \onlinecite{HuPRB2008} one can define a Faraday rotation angle experienced by a linearly polarised probe in transmission (rather than in reflection configuration), namely: $ \varphi _F^ \downarrow   = \frac{{\varphi _{cav}  - \varphi _{dot - cav} }}{2}$, where $\varphi _{cav}$ is the phase shift acquired by the linearly polarised beam in a cavity without an embedded dot, and $\varphi _{dot - cav}$ is the phase shift of the cavity-dot system. Using this definition, the Faraday rotation angle inferred from the calculated phase shift upon circularly polarised $\sigma^{-}$ excitation is $\approx \pm \pi/4$.

Although our results for the phase shifts are computed in the weak-coupling regime, they are very similar to the phase shift calculated in \cite{HuPRB2008} in the strong-coupling regime. The approach adopted in the latter is based on an approximate solution of the Heisenberg equations of motion for the cavity field ($\hat a$) and the dipole operator of the negative trion fundamental transition ($\hat \sigma^{-}$). Within this approach, the complex reflection coefficient of the dot-cavity system is obtained in the steady state and assuming that the trion state is most of the time in its ground state. Note that the shape of the phase shift calculated by our method, shown in Fig.~\ref{fig:phase_shift_circ}(b) is the same as the one calculated in \cite{HuPRB2008} shown in Fig.~2(b) (dotted curve) for a coupled dot-cavity, referred to as 'hot cavity'. We attribute this similarity to the coherent regime that we are working in. The driving high-intensity ultrashort pulse with pulse duration $T_p \ll \kappa, \Gamma_{\tau}$ leads to coherent propagation effects (e.g. Self-Induced Transparency) and polariton formation even without a cavity. The cavity may operate in the weak coupling regime, however the light-matter coupling of such a high-intensity pulse is sufficient to form a polaritonic travelling wave.

In order to investigate how this phase shift changes with the number of DBR pairs, we have performed simulations on another two structures, containing $22(5)$ and $25(5)$ pairs of layers. The large phase shift in the weak-coupling regime is confirmed by the calculated phase shift spectra displayed in Fig.~\ref{fig:phase_shift_circ}(b,c). The phase shift can acquire a value of $\pi$ depending on the number of DBR pairs, leading to a change of sign (cf. Fig.~\ref{fig:phase_shift_circ}(a) and (b)). Note that the offset of the cavity-dot phase with respect to the bare QD one remains the same independently of the number of layers. In contrast to the weak-coupling regime, two phase features (resonances) corresponding to the new polariton (dressed states) in the strong-coupling regime appear \cite{HuPRB2008} and the total polarisation rotation angle calculated is $\pm \pi/2$.

Due to the symmetry of the fundamental singlet trion energy-level structure and under the assumption of equality of the spin-flip population transfer rates $\gamma_{13}= \gamma_{31}=\gamma_1$ and $\gamma_{24}= \gamma_{42}=\gamma_3$ , a $\sigma{^+}$ excitation of an initially prepared in spin-up ground state would lead to the same dynamics and shift. However, we expect the dynamics to be different for a $\sigma{^-}$ excitation of ground state $|3\rangle$ initially prepared in spin-up, as previously demonstrated for a single QD without a cavity \cite{SlavchevaPRB2008}.

\subsection{Quantum evolution upon linearly polarised excitation}
\label{sssec:3.3}
\begin{figure}
 \resizebox{.9\textwidth}{!}{%
\includegraphics[height=5 cm]{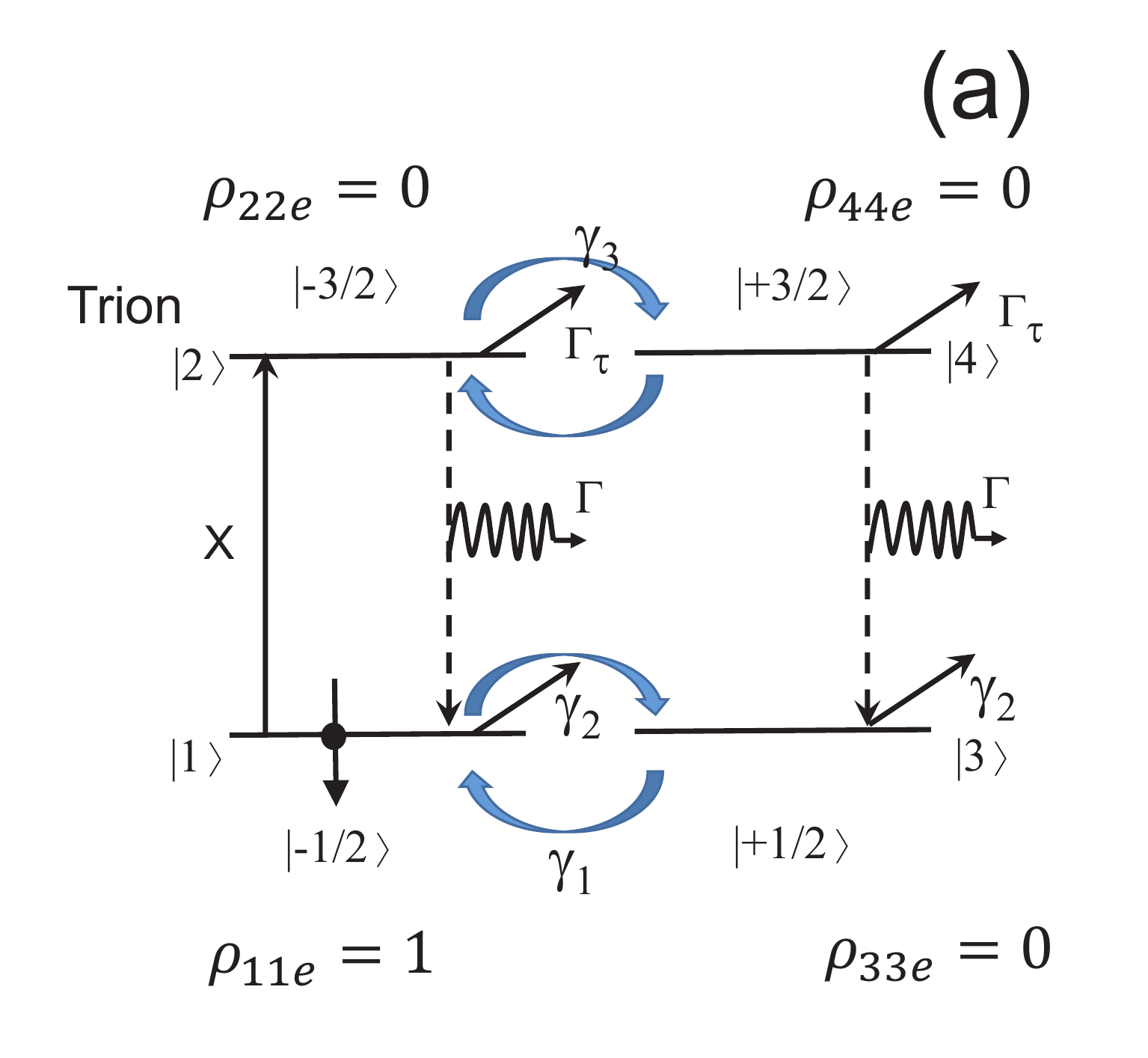}
\quad
\includegraphics[height=5 cm]{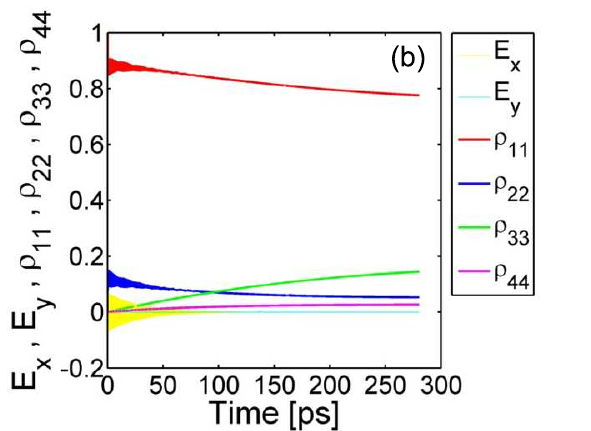}
\quad
\includegraphics[height=5 cm]{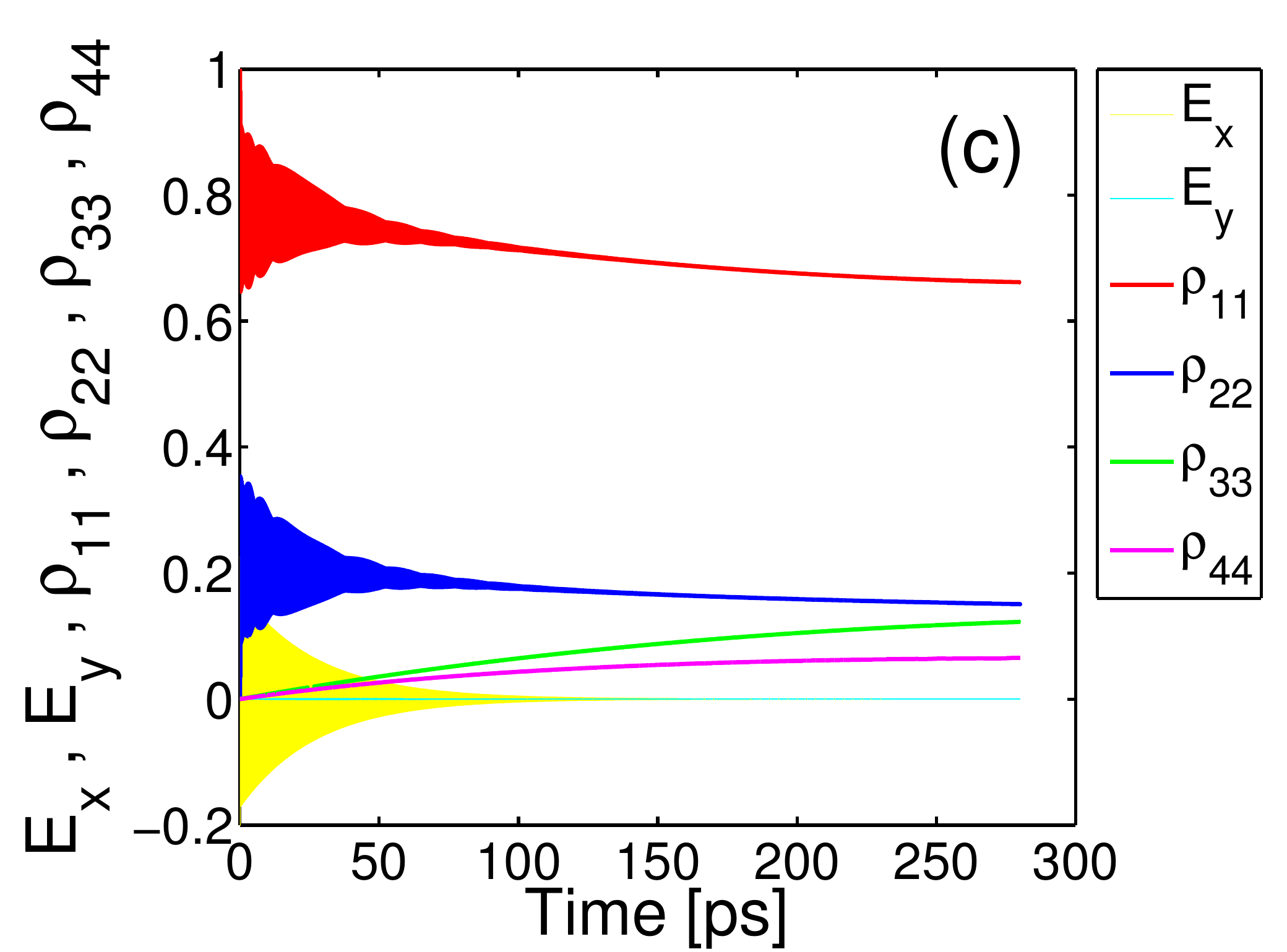}
}
 \resizebox{.9\textwidth}{!}{%
\includegraphics[height=5 cm]{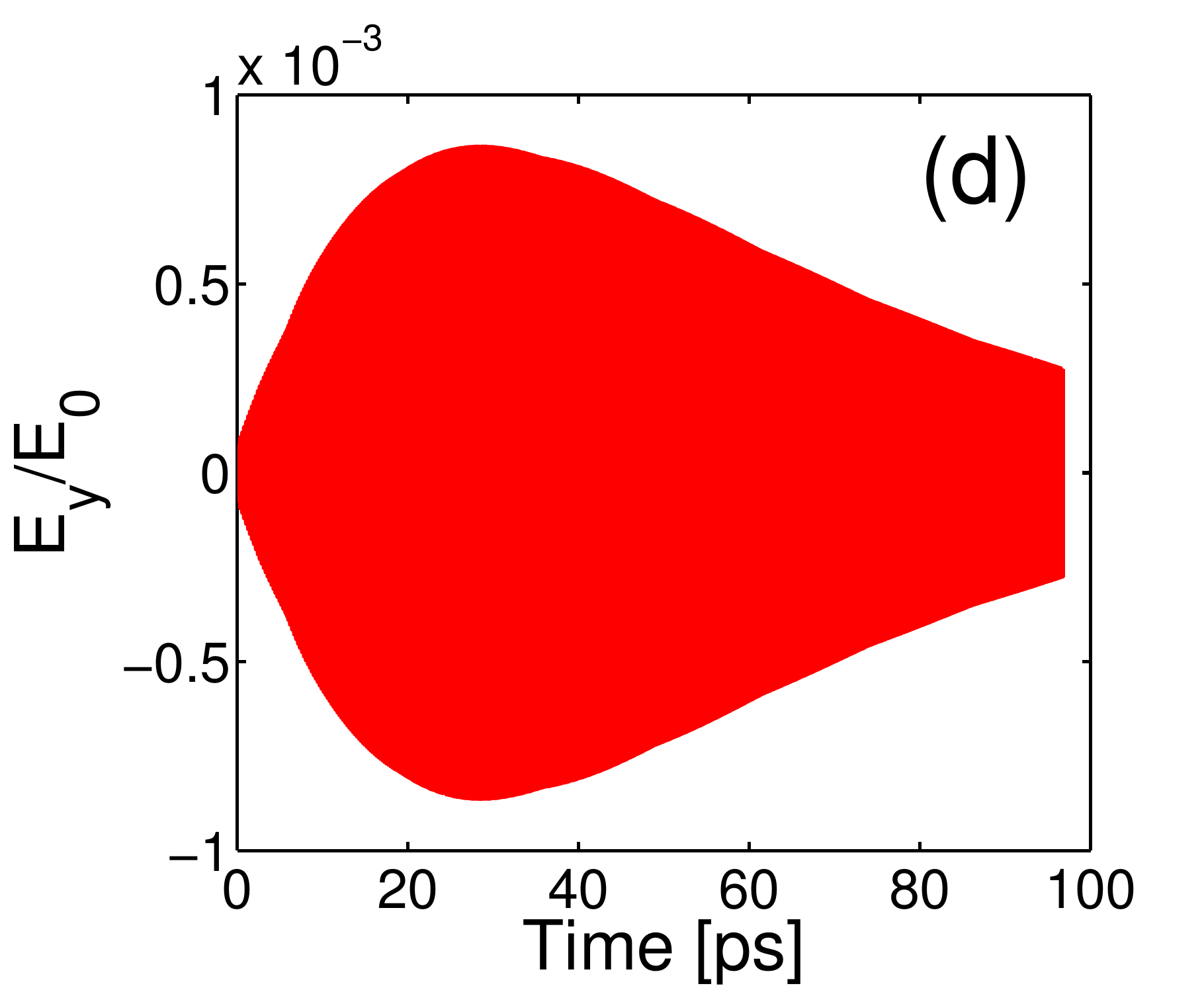}
\quad
\includegraphics[height=5 cm]{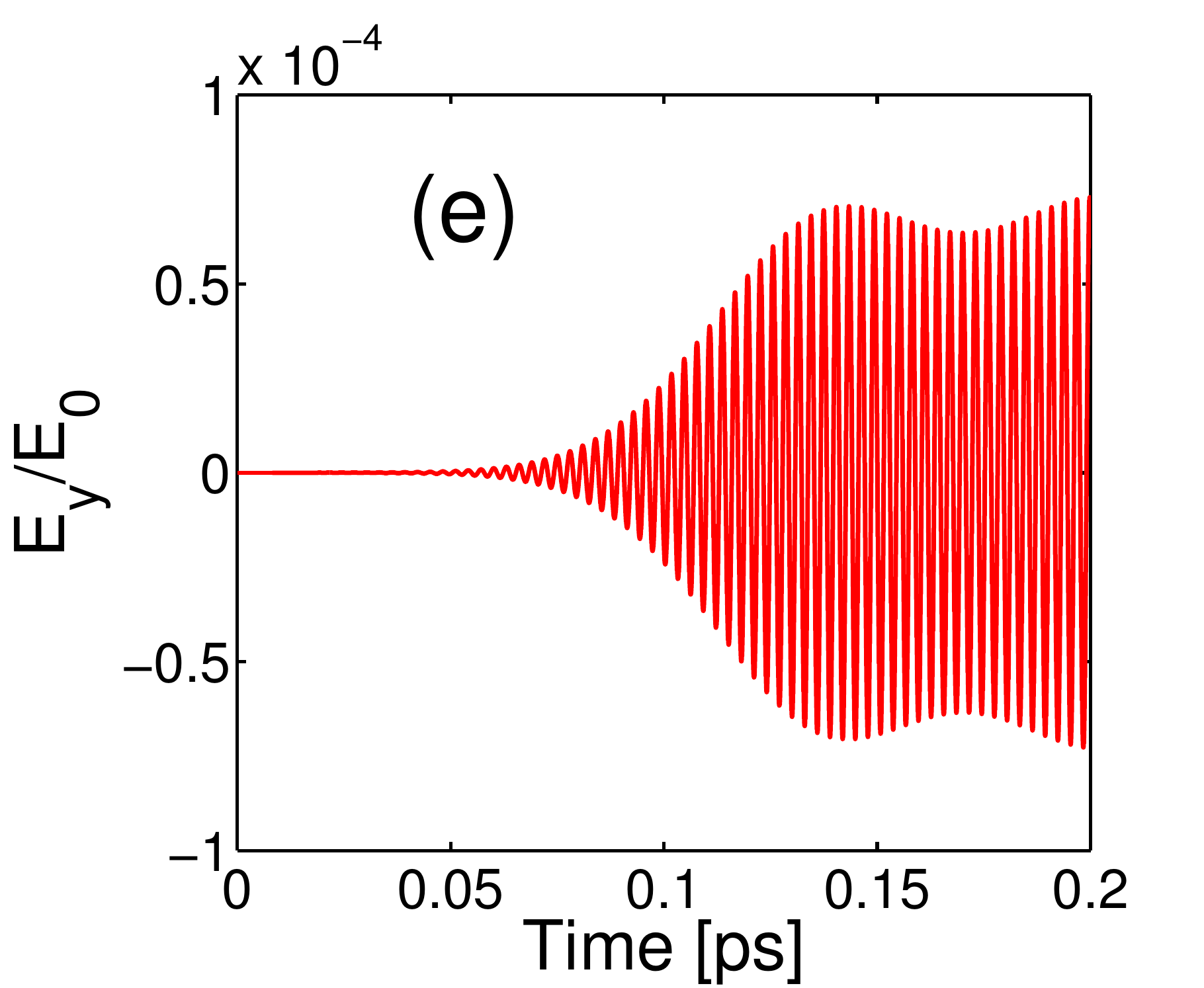}
}
\caption{(a) Excitation scheme with $x$-linearly polarised pulse and a system initially prepared in spin-down state; Time evolution of the normalised electric-field components of an $x$-linearly polarised pulse and level populations at the (b) left ($z_1$) and (c) right ($z_2$) ends of the QD layer (Fig.~\ref{fig:geometry}(b)). The population of level $|2\rangle$ (blue curve) is proportional to the experimentally detected time-resolved photoluminescence; (d) Zoom-in of (b,c) showing build-up of the $E_y$ electric field component in time from initial value zero to a maximum amplitude of $\sim 10^{-3} E_0$ and a subsequent amplitude decrease; (e) Zoom-in of the initial time evolution of (d) showing the ultrashort-time dynamics of the $E_y$ component amplitude build-up at $z=z_2$.}
\label{fig:linear_polaris}
\end{figure}

We now consider $x$-linearly polarised optical excitation of the quantum system described in Fig.~\ref{fig:linear_polaris}(a) in which the initial spin population is prepared in spin-down state. The source field is given by Eq.~\ref{eq:lin_polarised_pulse} where the $E_y$ field component is initially set to zero.
The time evolution of the electric field and the populations of all four levels at $z_1$ and $z_2$ upon $x$-linearly polarised pulse is shown in Fig.~\ref{fig:linear_polaris}(b,c) for a quantum system initially prepared in spin-down state (level $|1\rangle$). When zooming in the electric field components in Fig.~\ref{fig:linear_polaris}(b,c), a build-up of the $E_y$-component in time is revealed (see Fig.~\ref{fig:linear_polaris}(d)). This effectively means that the polarisation plane of the linearly polarised optical pulse is rotated during the pulse propagation. The maximum rotation angle in this case can be obtained from Fig.~\ref{fig:linear_polaris}(d) by $ \varphi _{\max }  = \arctan \left( {{\raise0.7ex\hbox{${E_y }$} \!\mathord{\left/ {\vphantom {{E_y } {E_x }}}\right.\kern-\nulldelimiterspace} \!\lower0.7ex\hbox{${E_x }$}}} \right) \approx 0.3\,^ \circ$.

\begin{figure}
 \resizebox{.9\textwidth}{!}{%
\includegraphics[height=5 cm]{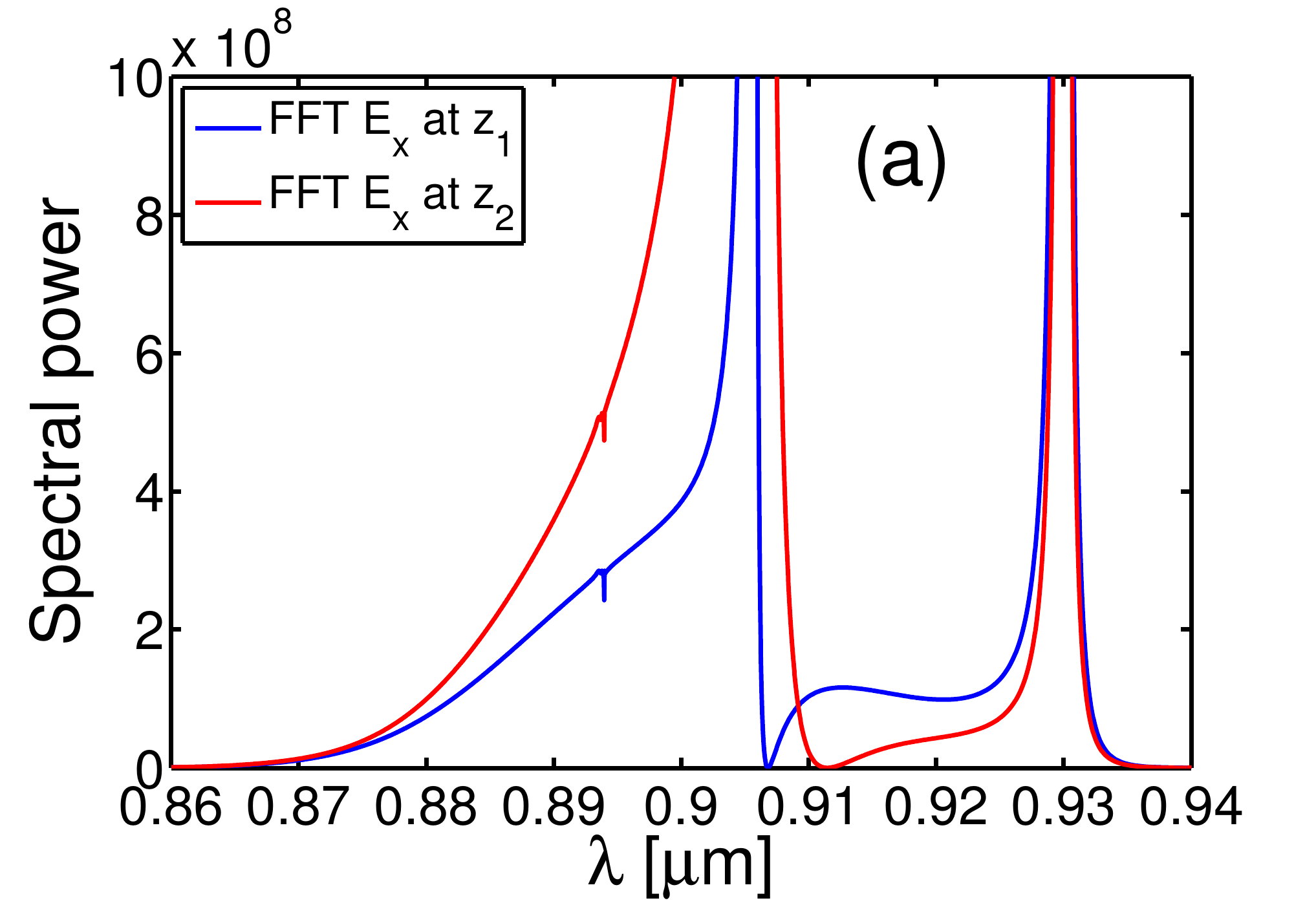}
\quad
\includegraphics[height=5 cm]{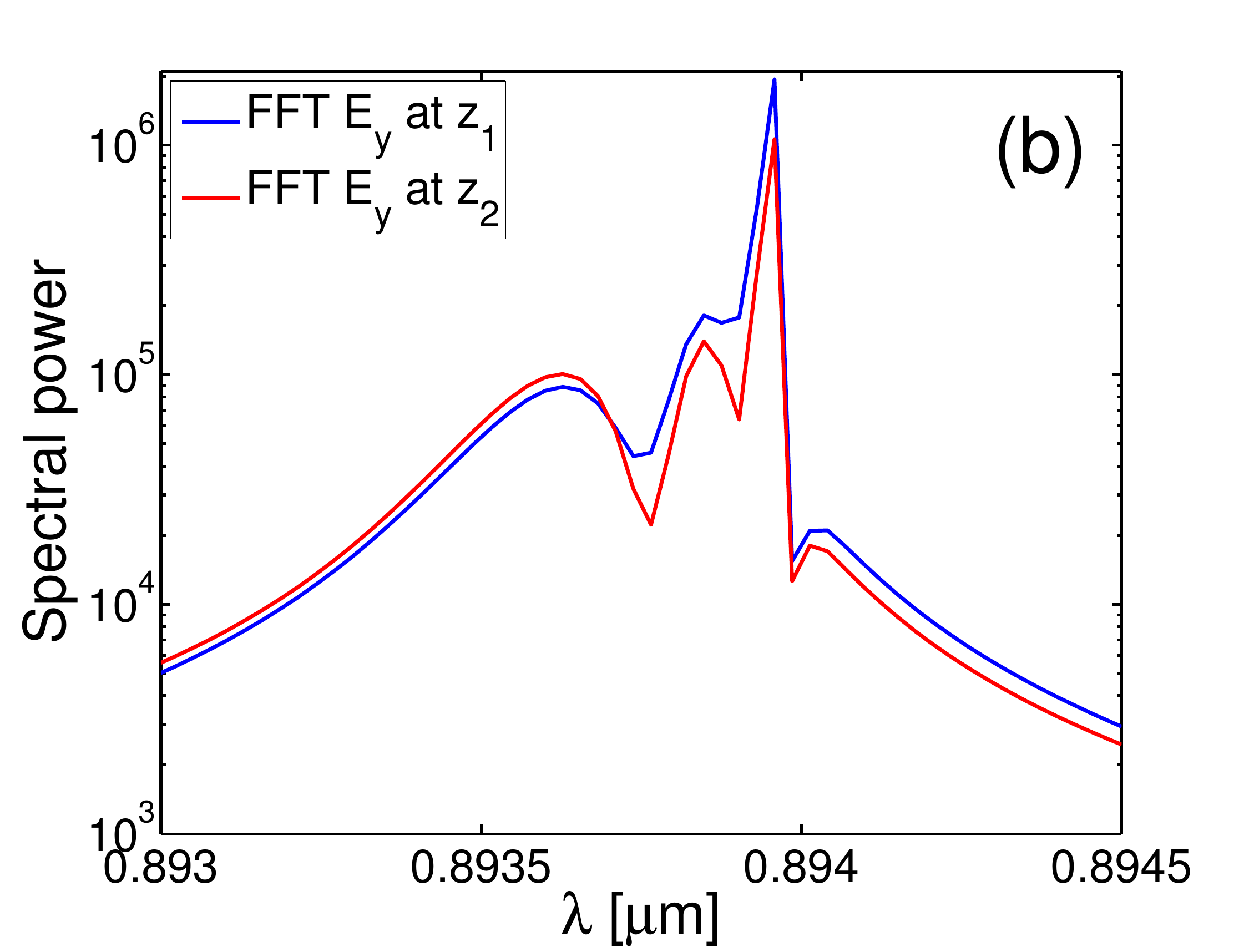}
}
\caption{ (a) Fourier (not normalised) transmission spectrum of the $E_x$ field component at $z_1$ (blue) and $z_2$ (red curve) as a function of wavelength (the amplitude squared of the discrete Fourier (FFT) transform is plotted). The dip observed at the resonant wavelength, $\lambda_0=894 \,\,\mathrm{nm}$, corresponds to resonant absorption of the pulse energy by the QD trion transition; (b)  Fourier spectrum of the $E_y$ field component at $z_1$ (blue) and $z_2$ (red curve) in semilogarithmic scale as a function of the pulse wavelength showing a peak at the trion transition resonance wavelength $\lambda_0=894 \,\,\mathrm{nm}$ with satellite lobes. While the $E_x$ component is resonantly absorbed (a), the $E_y$ field component amplitude is resonantly amplified (b).}
\label{fig:Fourier_lin_polar}
\end{figure}

Following the procedure for phase shift calculation described in \ref{sssec:3.2} we calculate the Fourier spectra of the time traces detected at the left ($z_1$) and right ($z_2$) ends of the active QD layer. The Fourier spectra of the $E_x$ field component at $z_1$ and $z_2$ are shown in Fig.~\ref{fig:Fourier_lin_polar} (a) and the ones corresponding to the orthogonal $E_y$ component is shown in (b) on a semilogarithmic scale. The sharp dip feature in (a) corresponds to the resonant absorption at the QD trion resonance wavelength $\lambda_0=894 \,\,\mathrm{nm}$. By contrast, Fig.~\ref{fig:Fourier_lin_polar}(b) shows a peak in the Fourier spectra at the trion resonance wavelength which is a signature of resonant amplification of the $E_y$ component. Thus while the $E_x$ component is resonantly absorbed, the $E_y$ pulse component is resonantly amplified, resulting in the amplitude build-up over time shown in Fig.~\ref{fig:linear_polaris}(d). We attribute the satellite peaks in Fig.~\ref{fig:Fourier_lin_polar} to constructive interference effects within the micropillar cavity.

\begin{figure}
 \resizebox{.9\textwidth}{!}{%
\includegraphics[height=5 cm]{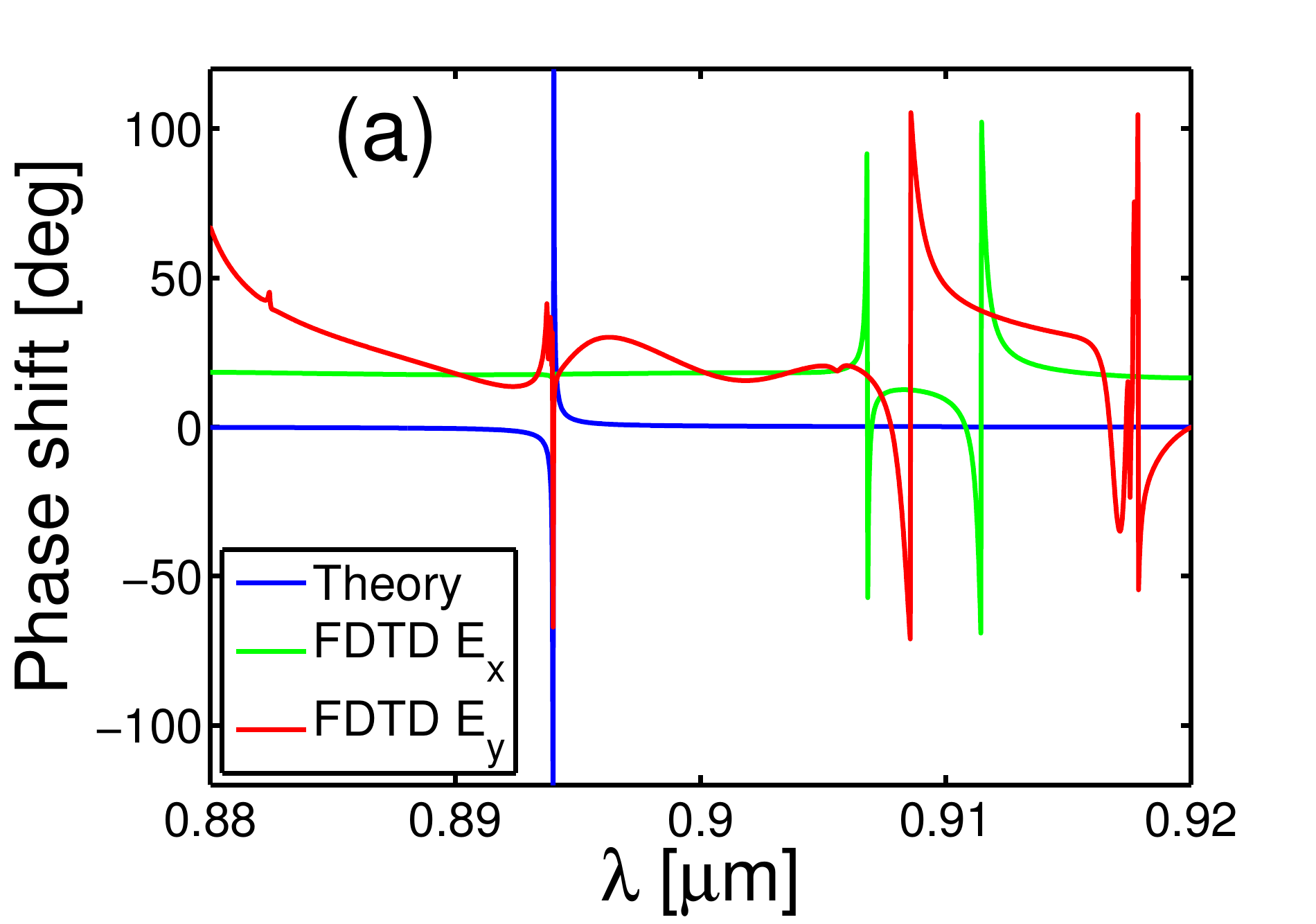}
\quad
\includegraphics[height=5 cm]{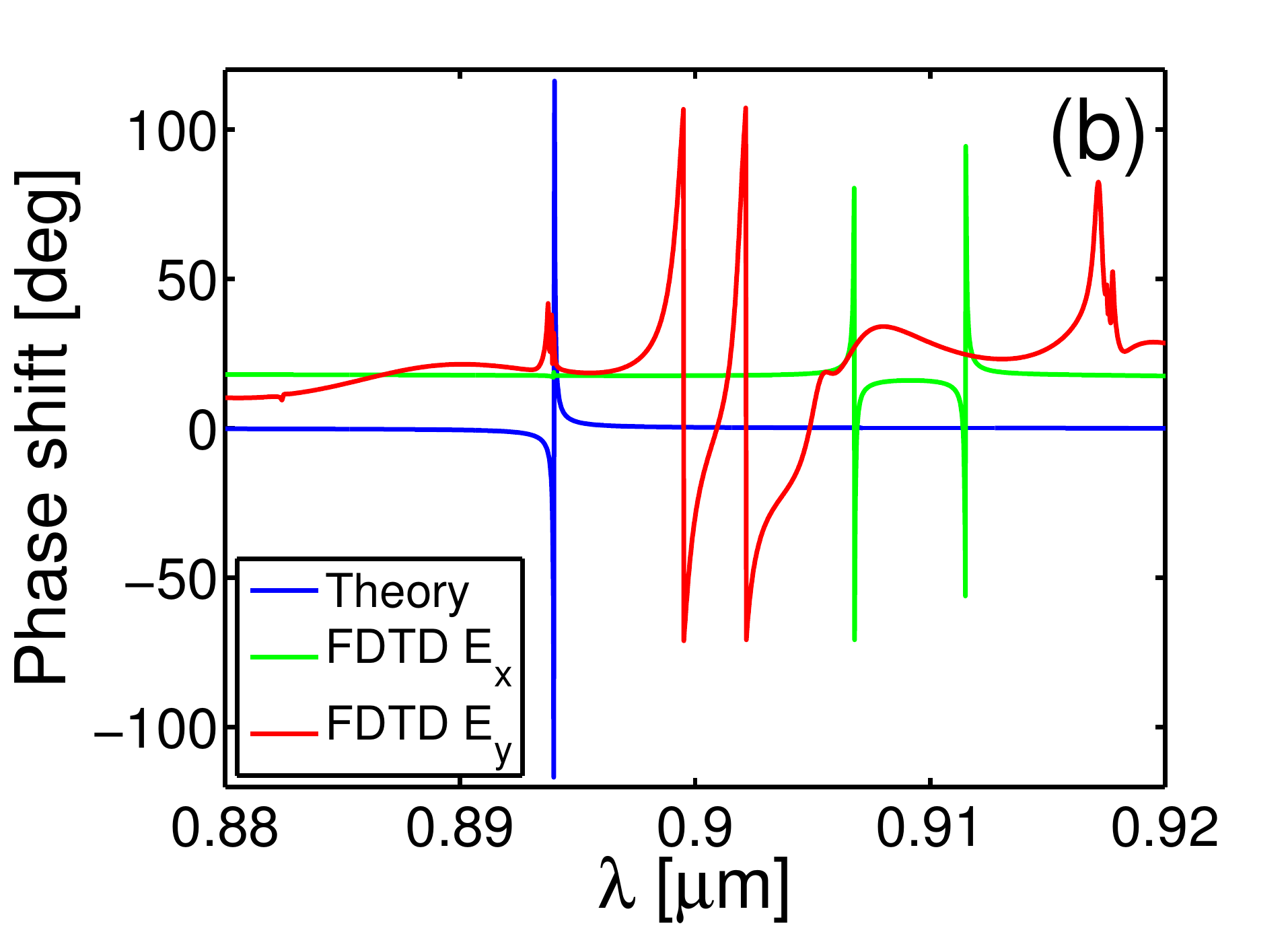}
}
\caption{(a) Phase shift (in degrees) vs wavelength of $E_x$ (green) and $E_y$ (red) field components of an $x$-linearly polarised pulse tuned in resonance with the QD trion transition (the phase shift of a homogeneously broadened two-level system at resonance is shown in blue as a reference). The $E_y$ component phase shift is red-shifted with respect to the $E_x$ one, resulting in splitting between the two spectra and an effective reduction in the detected optical polarisation rotation angle in this polarisation; (b) Phase shift spectrum of the $E_x$ and $E_y$ components vs wavelength for an $x$-linearly polarised pulse resonantly exciting the QD ground trion transition, which is initially in thermal equilibrium (see Fig.~ \ref{fig:lin_polaris_half_half}(a)). The $E_y$ spectrum is blue-shifted with respect to the $E_x$ spectrum, resulting effectively in a rotation angle reduction.}
\label{fig:lin_polaris}
\end{figure}

From the Fourier transmission spectra, we calculate the complex propagation factor and the phase shift induced by the resonant QD trion transition. The calculated phase shift for the $E_x$ and $E_y$ field components is displayed in Fig.~\ref{fig:lin_polaris}(a). Note that contrary to the circularly polarised excitation considered in \ref{sssec:3.1}, where the phase shift spectra of both components coincide (Fig.~\ref{fig:phase_shift_circ}), the phase shift spectrum of the $E_y$ field component is red-shifted with respect to the $E_x$ field component. Clearly, there is a correspondence between the time domain and the frequency domain. The red shift of the $E_y$ phase shift spectra vs wavelength with respect to the $E_x$ one may be attributed to the time delay  ($\approx 30 \,\mathrm{ps}$,see Fig. \ref{fig:linear_polaris}(d)) with which the $E_y$ component resonantly builds up within the cavity in the time domain. The two elliptically polarised pulse components oscillate with different frequencies. Therefore, the beatings between the two components may result in a destructive interference in the detection wavelength range, leading to a significant reduction of the experimentally observable optical rotation angle (as confirmed by our calculation of $\phi_{max}$ above). For instance, much lower rotation angles ($\sim 6^{\circ}$) have been reported for linearly polarised excitation. Hence our calculations of a realistic micropillar-dot structure show that the optical rotation angle of a circularly polarised pulse is much larger than the one for a linearly polarised one.

\subsection{Quantum evolution upon linearly polarised excitation of a quantum system initially in thermal equilibrium}
\label{sssec:3.4}
\begin{figure}
 \resizebox{.9\textwidth}{!}{%
\includegraphics[height=5 cm]{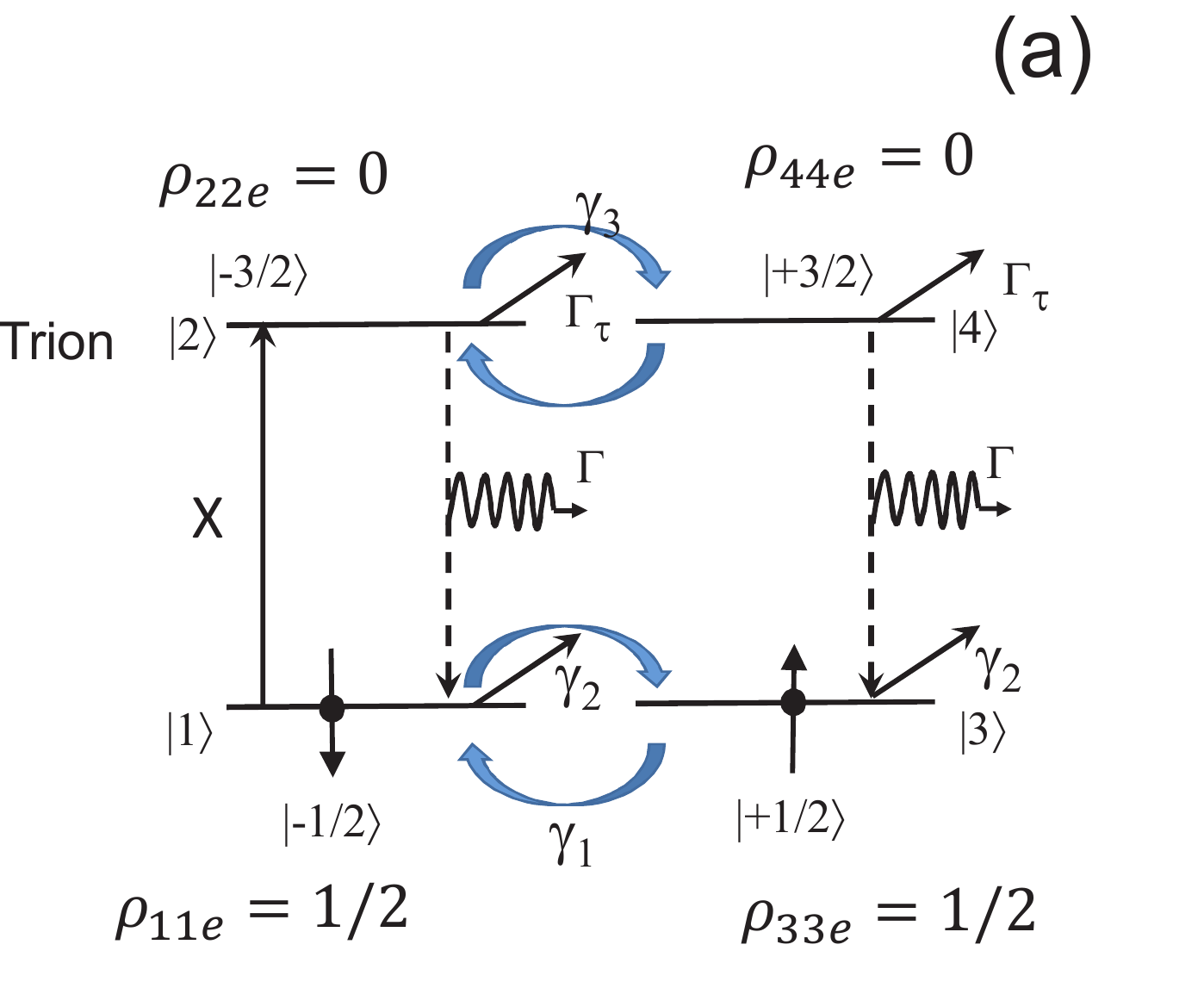}
\quad
\includegraphics[height=5 cm]{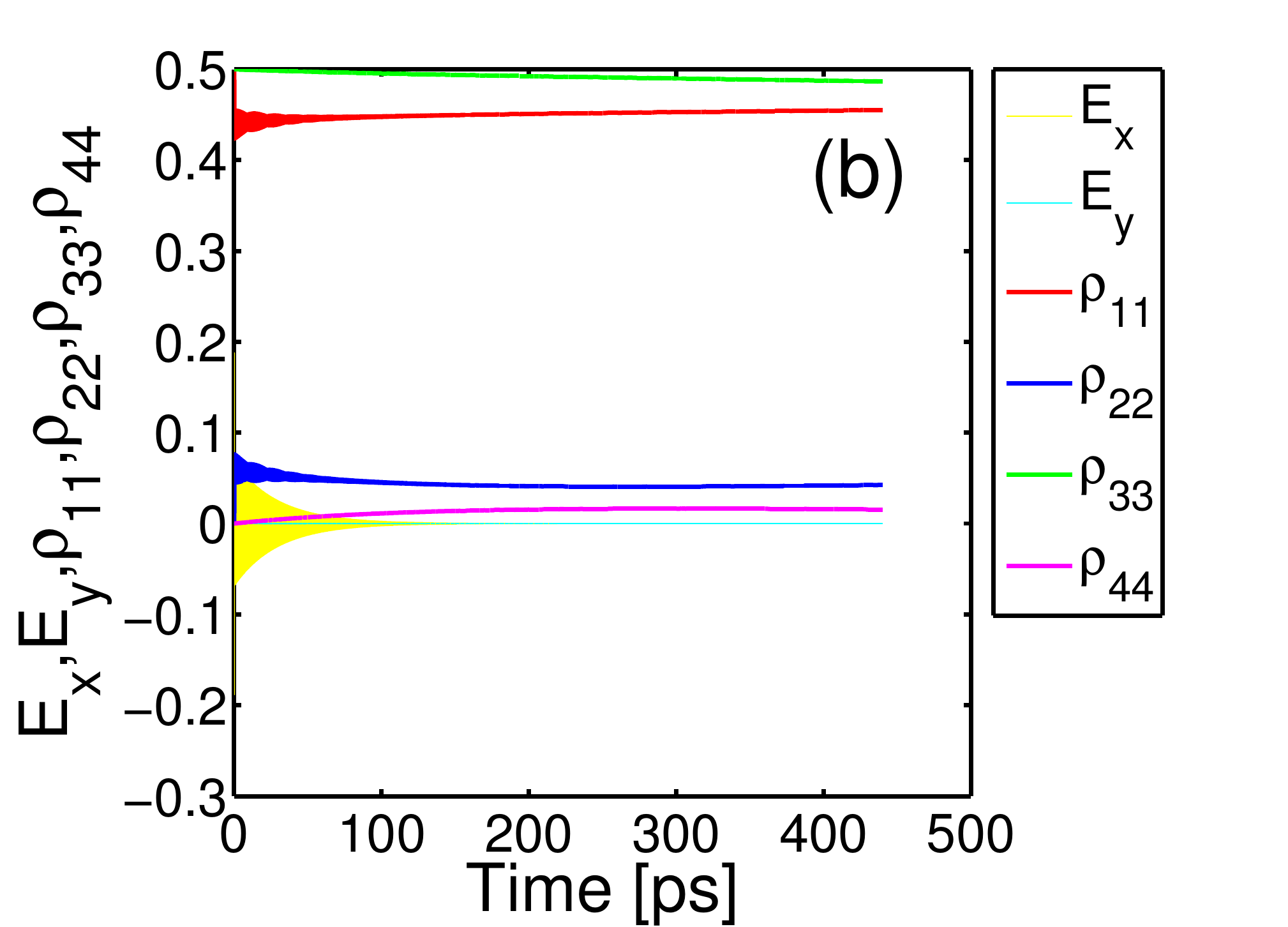}
\quad
\includegraphics[height=5 cm]{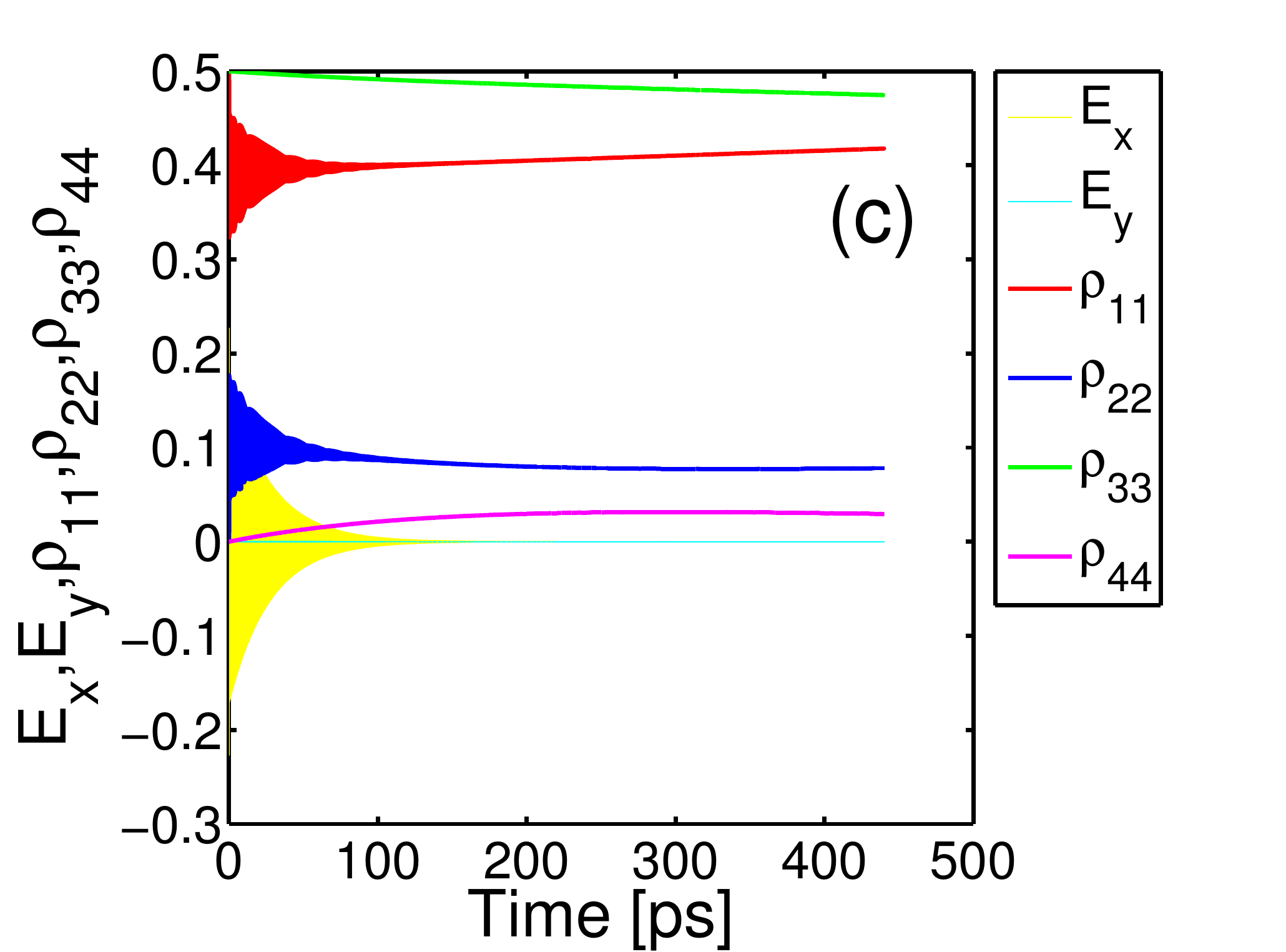}
}
\caption{(a) Negative trion energy-level scheme under $x$-linearly polarised pulse excitation; initial boundary conditions: spin-down and spin-up populations in thermal equilibrium,i.e. equally distributed between level $|1\rangle$ and $|3\rangle$; Time evolution of $E_x,E_y$ field components and spin populations of all four levels at (b) $z=z_1$; (c) $z=z_2$.}
\label{fig:lin_polaris_half_half}
\end{figure}

Finally, we consider the case of a linearly polarised pulse exciting a QD trion transition which is initially in thermal equilibrium, so that the spin-up and spin-down populations of the doubly-degenerate ground level in Fig.~\ref{fig:lin_polaris_half_half}(a) are equally distributed between the ground levels ($|1\rangle$ and $|3\rangle$). The computed time traces of the field components ($E_y=0$ initially) and the spin populations of all four levels at the left ($z_1$) and right ($z_2$) ends of the QD active layer are plotted in Fig.~\ref{fig:lin_polaris_half_half}(b,c), respectively.

The Fourier spectra of the $E_x$ field components at $z_1$ and $z_2$ (not shown) exhibit less pronounced transmission dips, a signature of the pulse resonant absorption at the QD trion transition, compared to Fig.~\ref{fig:Fourier_lin_polar}(a) and the $E_y$ component transmission is enhanced to a lesser degree. The induced phase shift is displayed in Fig.~\ref{fig:lin_polaris}(b). Note that the $E_y$ phase shift in this particular case is blue-shifted with respect to the $E_x$ spectrum and the $E_y$ component oscillates once again with a different frequency, effectively resulting in a destructive interference and a rotation angle reduction. The shape of the phase shift curve is similar to the one in Ref. \onlinecite{HuPRB2008} (Fig. 2 (b) -dotted curve corresponding to a 'hot cavity') and the two phase features are closely spaced. This behaviour is due to the complex spin population dynamics, which involves two simultaneous transitions from both ground levels excited by the left and right circularly polarised components of the linearly polarised pulse. The blue shift and the closer spacing reflects the time dynamics which from the very beginning includes both recombination channels, leading to a faster accumulation of the $E_y$ component phase compared to the $E_x$ one. In contrast to the assumptions in \onlinecite{HuPRB2008}, it is no longer possible to consider the system mostly in the ground state. This result demonstrates the importance of the initial conditions for the quantum evolution of the system, which in turn determines the angle of polarisation rotation.

By comparing Fig.~\ref{fig:lin_polaris}(a) and (b) upon a linearly polarised excitation, we note that the $E_y$ field component of a four-level system initially prepared in spin-down state (a) is red-shifted with respect to the $E_x$ component one, while $E_y$ is blue-shifted in the case of equally distributed spin-up and spin-down populations (b). This implies that the experimentally detected phase shift (or Faraday rotation angle) with respect to the $E_x$ component will be in an opposite direction for both cases, which in turn allows to distinguish between the two cases of initial ground state spin population preparation.

\section{Conclusions}
We have developed a dynamical model of a realistic open coupled cavity-dot system -- a negatively charged QD embedded in a micropillar cavity, upon an ultrashort optical excitation resonantly interacting with the QD trion transition. Unlike previous theories, our model treats simultaneously both the short and long time optically-induced spin dynamics and the incoherent spin relaxation and decoherence processes are fully taken into account. The temporal and spatial dynamics of the cavity-dot system are numerically computed and the phase shift induced by the confined single spin on the ultrashort optical pulse is inferred from the complex propagation factor in the active QD layer.

We demonstrate numerically a giant $\sim \pm \pi/2$ phase shift for a circularly polarised ultrashort pulse, which exceeds by an order of magnitude the Kerr polarisation rotation angles obtained by cw linearly polarised excitations. Our results point out to considerably lower polarisation rotation angles under linearly polarised and/or cw excitation. In addition, our computations shed light on the importance of the initial preparation of the system in a particular spin state. We show that maximum rotation angle is achieved for a system initially prepared in either spin-down or spin-up state (due to symmetry of the ground singlet trion energy level system).

Realisation of spin-photon entangler is a route for enabling gate operations with photons. We have numerically demonstrated that the cavity-dot structures are suitable candidates for realisation of controlled-phase gate functionalities on a chip for a next generation photon-based quantum logic. On the other hand, the cavity-dot systems could perform the function of ultrafast and reliable polarisation switches based on the Faraday rotation effect.

Our method allows to design and test cavity-dot structures with a view of maximising the photon polarisation rotation angle and thus prepare high-fidelity photon polarisation states for integrated quantum photonics applications. In addition, such large rotation angles would allow reliable detection of the initial spin state which will relax the requirements for highly sensitive polarisation detectors on a chip. For instance, one could envisage using cavity-dot structures in an interferometer configuration for measuring the phase shifts. The circular dichroism and birefringence of a charged dot-cavity system without and under an external magnetic field will be a subject of a future study.
\appendix
\begin{subappendices}
\section{Spin relaxation and decoherence in a four-level system}
The longitudinal spin relaxation processes in this particular level configuration is described by a sparse $16 \times 16$ block-diagonal decay rates matrix, given by:
\begin{equation}
\mathord{\buildrel{\lower3pt\hbox{$\scriptscriptstyle\frown$}}
\over \Gamma } _{damp}  = \sum\limits_{k = 1}^4 {\sum\limits_{i = 1}^4 {\left( {\Gamma _{ik}  - \Gamma _{ki} } \right)\left( {1 - \delta _{ik} } \right)} }
\end{equation}
where $\Gamma_{ik},\,\ i,k=1,...,4$ are the spin population transfer rates between each pair of levels. We denote the diagonal elements of the above block matrix by $\Gamma_i, \,\ i=1,...,4$, each being a $4\times4$  matrix, describing the spin relaxation of the diagonal (population) components of the density matrix. The matrices are given explicitly by:
\begin{equation}
\Gamma _1  = \left( {\begin{array}{*{20}c}
   { - \gamma _{13} } & 0 & 0 & 0  \\
   0 & \Gamma  & 0 & 0  \\
   0 & 0 & {\gamma _{31} } & 0  \\
   0 & 0 & 0 & 0  \\
\end{array}} \right),\Gamma _2  = \left( {\begin{array}{*{20}c}
   0 & 0 & 0 & 0  \\
   0 & { - \Gamma  - \gamma _{24} } & 0 & 0  \\
   0 & 0 & 0 & 0  \\
   0 & 0 & 0 & {\gamma _{42} }  \\
\end{array}} \right),\Gamma _3  = \left( {\begin{array}{*{20}c}
   {\gamma _{13} } & 0 & 0 & 0  \\
   0 & 0 & 0 & 0  \\
   0 & 0 & { - \gamma _{31} } & 0  \\
   0 & 0 & 0 & \Gamma   \\
\end{array}} \right),\Gamma _4  = \left( {\begin{array}{*{20}c}
   0 & 0 & 0 & 0  \\
   0 & {\gamma _{24} } & 0 & 0  \\
   0 & 0 & 0 & 0  \\
   0 & 0 & 0 & { - \Gamma  - \gamma _{42} }  \\
\end{array}} \right)
\end{equation}
where we have used the notations for the ralaxation rates in Fig.~\ref{fig:dot_micropillar}(b), setting
\begin{equation}
\Gamma _{12}  = \Gamma _{21}  = \Gamma _{34}  = \Gamma _{43}  = \Gamma ,\,\,\,\Gamma _{13}  = \gamma _{13} ,\,\,\Gamma _{31}  = \gamma _{31} ,\,\,\,\Gamma _{24}  = \gamma _{24} ,\,\,\Gamma _{42}  = \gamma _{42}
\end{equation}
We have shown \cite{SlavchevaPRB2008} that the longitudinal relaxation term in (\ref{eq:Liouville_master}) is given by:
\begin{equation}
\mathord{\buildrel{\lower3pt\hbox{$\scriptscriptstyle\frown$}}
\over \sigma }  = diag\left( {Tr\left( {\mathord{\buildrel{\lower3pt\hbox{$\scriptscriptstyle\frown$}}
\over \rho.\mathord{\buildrel{\lower3pt\hbox{$\scriptscriptstyle\frown$}}
\over \Gamma } _i} } \right)} \right),\,\,\,i = 1,...,4
\end{equation}
The transverse relaxation rate matrix, $\hat \Gamma_t$ in  (\ref{eq:Liouville_master}) is given by the off-diagonal part of $\hat \sigma$, representing the relaxation of the dipole moments between each pair of levels within the system, due to spin decoherence processes involving levels outside the four-level system considered. The latter is a symmetric matrix, given in terms of Fig.~\ref{fig:dot_micropillar}(b) notations as:
\begin{equation}
\mathord{\buildrel{\lower3pt\hbox{$\scriptscriptstyle\frown$}}
\over \Gamma } _t  = \left( {\begin{array}{*{20}c}
   0 & {\Gamma _\tau  } & {\gamma _2 } & {\gamma _2 }  \\
   {\Gamma _\tau  } & 0 & {\Gamma _\tau  } & {\Gamma _\tau  }  \\
   {\gamma _2 } & {\Gamma _\tau  } & 0 & {\Gamma _\tau  }  \\
   {\gamma _2 } & {\Gamma _\tau  } & {\Gamma _\tau  } & 0  \\
\end{array}} \right)
\end{equation}

\label{sec:1}

\end{subappendices}

\end{document}